\documentstyle[12pt,psfig]{article}
\def\gsim{\stackrel{>}{{}_\sim}}
\def\lsim{\stackrel{<}{{}_\sim}}

\def\simgt{\stackrel{>}{{}_\sim}}
\def\be{\begin{equation}}
\def\ee{\end{equation}}
\def\bear{\be\begin{array}}
\def\eear{\end{array}\ee}
\def\bea{\begin{eqnarray}}
\def\eea{\end{eqnarray}}

\def\ie{{\it i.e.}}
\def\etal{{\it et al.}}
\def\half{{\textstyle{1 \over 2}}}
\def\third{{\textstyle{1 \over 3}}}

\def\sixth{{\textstyle{1 \over 6}}}
\def\eighth{{\textstyle{1 \over 8}}}
\def\bold#1{\setbox0=\hbox{$#1$}%
     \kern-.025em\copy0\kern-\wd0
     \kern.05em\copy0\kern-\wd0
     \kern-.025em\raise.0433em\box0 }
\begin{document}
\catcode`@=11
\newtoks\@stequation
\def\subequations{\refstepcounter{equation}%
\edef\@savedequation{\the\c@equation}%
  \@stequation=\expandafter{\theequation}
  \edef\@savedtheequation{\the\@stequation}
  \edef\oldtheequation{\theequation}%
  \setcounter{equation}{0}%
  \def\theequation{\oldtheequation\alph{equation}}}
\def\endsubequations{\setcounter{equation}{\@savedequation}%
  \@stequation=\expandafter{\@savedtheequation}%
  \edef\theequation{\the\@stequation}\global\@ignoretrue

\noindent}
\catcode`@=12
\begin{titlepage}

\begin{flushright}
OUTP--95--36P\\
SHEP--95--35\\
IEM--FT--121/95 \\
November 1995
\end{flushright}

\vspace*{5mm}

\begin{center}
{\Huge Selectron Pair Production in $ e^+e^-$}\\
\vspace*{5mm}
{\Huge Colliders and the}\\
\vspace*{5mm}
{\Huge  Supergravity Spectrum}\\[15mm]

{\large{\bf Beatriz de Carlos ${}^{\dag}$}
and {\bf Marco A. D\'\i az ${}^{ {\ddag}}$}\\
\hspace{3cm}\\
${}^{\footnotesize\dag}$ {\small School of Mathematical and Physical
Sciences, University of Sussex} \\
{\small Falmer, Brighton BN1 9QH, U.K.}  \\
{\small e-mail: B.de-Carlos@sussex.ac.uk} \\
\vspace{0.3cm}
${}^{\footnotesize\ddag}$ {\small Physics Department, University of
Southampton }\\
{\small Southampton, SO17 1BJ, U.K.} \\
{\small e-mail: mad@hep.phys.soton.ac.uk}}
\end{center}
\vspace{5mm}
\begin{abstract}
Selectrons may be produced in pairs at LEPII if their mass
is less than about 100 GeV. Preferably, they decay into
the lightest neutralino plus an electron. In a scenario
where selectrons are observed at LEPII, we show that:\\
(i) in a first stage where experimental errors are large, the
measurement of the total cross section of selectron
pair production, the selectron mass, and the lightest
neutralino mass, allow us to validate or rule out the Minimal
Supergravity Model in its simplest form, and that\\
(ii) in a second stage where precision measurements are available,
the value of $\tan\beta$ can be determined together with the rest
of the parameters that specify the Minimal Supergravity Model and,
with them, the entire supersymmetric spectrum can be calculated.\\
We include experimental constraints from sparticle searches,
$Z$--pole physics, stability of the lightest supersymmetric particle
(LSP) and the decay $b\rightarrow  s\gamma$. In these scenarios, small
values of $\tan\beta$ and negative values of $\mu$ are preferred, and
the lightest Higgs mass satisfies $m_h<110$ GeV, which makes it likely
to be detected at LEPII.
\end{abstract}

\end{titlepage}

\setcounter{page}{1}

\section{Introduction}

Despite the success of the Standard Model (SM) in describing the
strong and electroweak interactions at energy scales accessible to
present colliders, this model is regarded as an incomplete theory.
Two of the main theoretical problems with the SM are related to the
electroweak symmetry breaking. First, it is hard to understand why
the scale of electroweak symmetry breaking is much smaller
compared to the Planck scale. Second, the existence of fundamental
scalars in the SM model is problematical since their masses are
unstable under radiative corrections. Both problems are potentially
solved in supersymmetric theories. For this and other reasons,
since its discovery \cite{susydisc}, supersymmetry has received
great attention and has become the leading candidate for physics
beyond the SM.

The simplest supersymmetric extension of the SM is called the
Minimal Supersymmetric Model (MSSM)
\cite{MSSMrep}\ and it is based
on R--parity conservation and minimal particle content.
Many experiments have looked for supersymmetric particles, but
no signal has been found so far \cite{PDG,L3Rep,AlephRep}.
The $e^+e^-$ annihilation is a clean environment for searching
for new particles, and the LEPII collider will have enough
energy to look for supersymmetric particles with mass up to
about half the center of mass energy,
since they are produced in pairs. Besides the
lightest Higgs boson, candidates to be found at LEPII are
charginos, neutralinos, and sleptons because they are typically
lighter than other susy particles. Among them,
the cleanest signatures are produced by charged particles:
charginos and charged sleptons.

If a supersymmetric particle is detected at LEPII, it is crucial to
know what can be learned about the model from the LEP measurements.
In the case of chargino pair production, this issue was addressed
recently in the case of global supersymmetry \cite{Feng,DiazKing}.
Furthermore, in supergravity (SUGRA) models with radiatively broken
electroweak symmetry, the predictions are more powerful
\cite{DiazKing2}. Nevertheless, charged sleptons can be lighter
than charginos, and for this reason a charged slepton may be the
first supersymmetric signal to be detected.

In supergravity models with radiatively broken electroweak symmetry,
selectrons are in general not lighter than staus and smuons and,
therefore, the latter may be detected first. But staus and smuons can
be produced in $e^+e^-$ colliders only with intermediate photons and
$Z$--bosons in the $s$--channel. Furthermore, couplings of charged
sleptons to $\gamma$ or $Z$ depend only on electroweak parameters like
the gauge coupling constant $g$, the weak mixing angle $\theta_W$, or
the electric charge $e$. This implies that other than the mass of the
stau or smuon, and the mass of the decay products (lightest
neutralino), it will be hard to extract more information on the
supersymmetric parameters of the model with the discovery of these
particles. This is not the case with selectrons. In addition to
intermediate $\gamma$ and $Z$--bosons, selectrons may be produced in
$e^+e^-$ colliders with intermediate neutralinos in the $t$--channel,
and therefore, its production cross section depends on the neutralino
sector and is sensible to important supersymmetric parameters.

In global supersymmetry, many authors have studied the selectron
pair production in $e^+e^-$ colliders \cite{pairseprod}, and also
the production of off-shell selectrons \cite{offshelse}. The
purpose of this paper is to study this process in the context
of supergravity models, where the parameters are tightly
constrained, and therefore the predictive power is greater.
We will demonstrate that the experimental observables associated
with the detection of a pair of selectrons, \ie, the total
production cross section, the mass of the selectron, and the
mass of its main decay product, the lightest neutralino, allow
us to calculate the supersymmetric parameters that define the model,
and through them, to predict the entire supersymmetric spectrum.

\section{Minimal Supergravity}

Minimal Supergravity is defined by a K\"ahler potential
$K=\sum_j|\phi_j|^2$ and a gauge kinetic function
$f_{ab}=\delta_{ab}$ so that all the kinetic terms are canonical.
The supersymmetric lagrangian is specified by the superpotential
$W$ given by\footnote{
We are using here the notation of ref.~\cite{GunHaber}. We advise
the reader that the sign of the $\mu$ term in the original version
of this paper was corrected in a subsequent erratum.
}
\begin{equation}
W=\varepsilon_{ab}\left[
 h_U^{ij}\widehat Q_i^a\widehat U_j\widehat H_2^b
+h_D^{ij}\widehat Q_i^b\widehat D_j\widehat H_1^a
+h_E^{ij}\widehat L_i^b\widehat R_j\widehat H_1^a\right]
-\mu\varepsilon_{ab}\widehat H_1^a\widehat H_2^b
\label{eq:Wsuppot}
\end{equation}
where $i,j=1,2,3$ are generation indices, $a,b=1,2$ are $SU(2)$
indices, and $\varepsilon$ is a completely antisymmetric $2\times2$
matrix, with $\varepsilon_{12}=-\varepsilon_{21}=1$ and
$\varepsilon_{11}=\varepsilon_{22}=0$. The symbol ``hat'' over each
letter indicates a superfield, with $\widehat Q_i$, $\widehat L_i$,
$\widehat H_1$, and $\widehat H_2$ being $SU(2)$ doublets with
hypercharges $\third$, $-1$, $-1$, and $1$ respectively, and $\widehat
U$, $\widehat D$, and $\widehat R$
being $SU(2)$ singlets with hypercharges $-{\textstyle{4\over 3}}$,
${\textstyle{2\over 3}}$, and $2$ respectively. The couplings $h_U$,
$h_D$ and $h_E$ are $3\times 3$ Yukawa matrices, and $\mu$ is
a parameter with units of mass.

In the supersymmetric part of the lagrangian, we find the Yukawa
interactions ${\cal L}_Y$ and the scalar potential $V_s$. In
${\cal L}_Y$ we get fermion mass terms and fermion-fermion-scalar
interactions, and is given by
\begin{equation}
{\cal L}_Y =- {1\over 2} \sum_{n,m} {{\partial^2\widetilde W}\over{
\partial A_n\partial A_m}}\psi_n\psi_m + {\rm h.c.}
\label{eq:Yuklag}
\end{equation}
where the function $\widetilde W$ is obtained by taking the
superpotential in eq.~(\ref{eq:Wsuppot})\ and replacing each
superfield by its scalar component. The indices $n$ and $m$ run over
all the superfields in the superpotential, and $A_n$ $(\psi_n)$ are
the scalar (fermionic) component of each superfield.

The scalar potential $V_s$ consists of two parts, the F terms and
the D terms
\begin{equation}
V_s=\sum_n F_n^*F_n+\half[D^aD^a+(D')^2] \; \; ;
\label{eq:Vscal}
\end{equation}
the F term is given by
\begin{equation}
F_n={{\partial\widetilde W}\over{\partial A_n}} \;\;,
\label{eq:Fterm}
\end{equation}
where $n$ runs over all the superfields in eq.~(\ref{eq:Wsuppot}),
and $A_n$ is the corresponding scalar component. The two D terms,
one for $SU(2)$ and one for $U(1)$, are
\begin{equation}
D^a=\half g\sum_n A_n^{i*}\sigma_{ij}^a A_n^j\,,\qquad
D'=\half g'\sum_m y_mA_m^*A_m\,.
\label{eq:Dterm}
\end{equation}
The first one is the $SU(2)$ D--term, where
$n$ runs over all the superfields which are doublets under
$SU(2)$, $A_n$ is the corresponding scalar component, $\sigma^a$
are the Pauli matrices, and $i,j=1,2$ are $SU(2)$ indices.
The second is the $U(1)$ D--term, where $m$ runs over all
the superfields with a non-zero hypercharge $y_m$, and
$A_m$ is the corresponding scalar component.

Supersymmetry must be broken because otherwise the known
fermions would be degenerate in mass with its superpartners
and this is not observed experimentally.
The actual supergravity mechanism is unknown
\footnote{For an example, see ref.~\cite{BdeCCM}.},
but can be parametrized with a set of soft supersymmetry
breaking terms which do not introduce quadratic divergences
to the unrenormalized theory \cite{softterms}
\begin{eqnarray}
V_{soft}&=&\varepsilon_{ab}\left[
 A_U^{ij}h_U^{ij}\widetilde Q_i^a\widetilde U_j H_2^b
+A_D^{ij}h_D^{ij}\widetilde Q_i^b\widetilde D_j H_1^a
+A_E^{ij}h_E^{ij}\widetilde L_i^b\widetilde R_j H_1^a\right]
-B\mu\varepsilon_{ab} H_1^a H_2^b\nonumber\\
&+&M_Q^{ij2}\widetilde Q^{a*}_i\widetilde Q^a_j+M_U^{ij2}
\widetilde U^*_i\widetilde U_j+M_D^{ij2}\widetilde D^*_i
\widetilde D_j+M_L^{ij2}\widetilde L^{a*}_i\widetilde L^a_j+
M_R^{ij2}\widetilde R^*_i\widetilde R_j \label{eq:Vsoft} \\
&+&m_1^2 H^{a*}_1 H^a_1+m_2^2 H^{a*}_2 H^a_2-
\left[\half M_s\lambda_s\lambda_s+\half M\lambda\lambda
+\half M'\lambda'\lambda'+h.c.\right]\,.\nonumber
\end{eqnarray}

In Minimal Supergravity the scalar masses, the gaugino masses,
and the trilinear couplings are universal at the unification
scale $M_X$
\begin{eqnarray}
&M_Q^{ij2}=M_U^{ij2}=M_D^{ij2}=M_L^{ij2}=M_R^{ij2}=m_0^2\delta^{ij}
\,,&\qquad m_1^2=m_2^2=m_0^2\,, \nonumber \\
&M_s=M=M'=M_{1/2}\,,&\qquad A_U^{ij}=A_D^{ij}=A_E^{ij}=A\delta^{ij}
\label{eq:universal}
\end{eqnarray}
and the mass parameters $A$ and $B$ are related to each other
at the scale $M_X$ by
\begin{equation}
A=B+m_0\,.
\label{eq:ABrel}
\end{equation}
In this case, only four parameters specify the model: the universal
scalar mass $m_0$, the universal gaugino mass $M_{1/2}$, the
universal trilinear coupling $A$, and the Higgs mass parameter $\mu$.

\subsection{The Effective Potential}

The Higgs sector of the MSSM \cite{HHG} contains two Higgs
doublets $H_1$ and $H_2$, with hypercharges $-1$ and $1$
respectively, and with the following tree level Higgs
potential:
\begin{eqnarray}
V_H&=&m_{1H}^2|H_1|^2+m_{2H}^2|H_2|^2-m_{12}^2
(\varepsilon_{ab}H_1^aH_2^b+h.c.)\nonumber\\
&+&\eighth(g^2+g'^2)\left[|H_1|^2-|H_2|^2\right]^2+
\half g^2|{H^a_1}^*H_2^a|^2\,,
\label{eq:Hpotential}
\end{eqnarray}
where $m_{1H}^2=m_1^2+\mu^2$, $m_{2H}^2=m_2^2+\mu^2$, and
$m_{12}^2=B\mu$. This potential has a minimum that breaks
the symmetry and the neutral component of the Higgs fields
get a vacuum expectation value:
\begin{equation}
H_1={{{1\over{\sqrt{2}}}[\chi^0_1+v_1+i\varphi^0_1]}\choose{
H^-_1}}\,,\qquad
H_2={{H^+_2}\choose{{1\over{\sqrt{2}}}[\chi^0_2+v_2+
i\varphi^0_2]}}\,,
\label{eq:shiftdoub}
\end{equation}
where $v_1^2+v_2^2=(246 \; {\rm GeV})^2$ is our normalization.

The one-loop effective potential \cite{effpot}, working in
dimensional reduction \cite{DRED}, is given by
\begin{equation}
V_{eff}=V_0(v_1,v_2,Q)+\sum_i{{w_i}\over{64\pi^2}}
m_i^4(v_1,v_2)\left(\ln{{m_i^2(v_1,v_2)}\over{Q^2}}-
{3\over 2}\right)\,.
\label{eq:effpot}
\end{equation}
The index $i$ runs over all the particles of the model,
$m_i(v_1,v_2)$ is the mass of the $i$-th particle as a
function of the two vacuum expectation values, and $w_i$
takes into account the internal degrees of freedom of
each particle: $w_i=(-1)^{2s}(2s+1)rc$, where $s$ is the spin
of the particle, $r=1(2)$ if the field is real (complex), and
$c$ is the color factor. The tree level effective potential
is found from eq.~(\ref{eq:Hpotential}) after replacing the
Higgs fields by their vacuum expectation values:
\begin{equation}
V_0(v_1,v_2,Q)={\textstyle{1\over 32}}(g^2+g'^2)(v_1^2-v_2^2)^2
+\half m_{1H}^2v_1^2+\half m_{2H}^2v_2^2-m_{12}^2v_1v_2\,.
\label{eq:V0eff}
\end{equation}
In $\overline{DR}$, one-loop contributions to the effective potential
are explicitly scale dependent and, therefore, the tree level
potential $V_0(v_1,v_2,Q)$ must be implicitly scale dependent, \ie,
each parameter in the right hand side of eq.~(\ref{eq:V0eff}) is
a running parameter governed by a renormalization group equation
(RGE).

The electroweak symmetry is broken radiatively because, despite the
fact that $m_1^2(Q=M_X)=m_2^2(Q=M_X)=m_0^2$, the large value of the
top quark mass makes the RGE of these two mass parameters to evolve
differently and $m_2^2(Q)$ is driven towards zero when the scale
approaches the weak scale. It was proven that, due to the strong scale
dependence of $V_0(v_1,v_2,Q)$, one-loop contributions to the
effective potential are important \cite{GaRiZw}, and to find reliable
results in the electroweak symmetry breaking, the one-loop
contributions from all the particles should be included
\cite{BdeCACfine}. One of the free parameters  must be fixed in order
to reproduce the correct value for the $Z$-boson mass. We choose to
fix $|\mu|$, therefore the model is specified by $m_0$, $M_{1/2}$,
$\tan\beta$, and  sign($\mu$) (we have eliminated $A$ in favor of
$\tan\beta$). In our calculations, we use the full one-loop effective
potential including the effect of all the particles in the model. We
minimize the potential at the scale  $\widehat Q$ defined by the scale
where the tree level vacuum expectation values are equal to the
one-loop corrected vev's. At this scale we find the running CP-odd
Higgs mass $m_A(\widehat Q)$\footnote{The scale $\widehat Q$ is the
only point where the running CP-odd Higgs mass is equal to its tree
level value $m_A^2(\widehat Q)=m_{1H}^2(\widehat Q)+ m_{2H}^2(\widehat
Q)$.}, which is related to the pole mass $m_A$ through the relation
\begin{equation}
m_A^2=m_A^2(\widehat Q)-(\delta
m_A^2)^{\overline{DR}}+ A_{AA}(p^2=m_A^2,\widehat Q) \;\; .
\label{eq:mApolerun}
\end{equation}
The function $A_{AA}(p^2,Q)$ is the CP-odd Higgs self energy evaluated
at an external momentum $p$ and at an arbitrary scale $Q$. This
function is divergent and its contribution from loops involving top
and bottom quarks and squarks is given in the appendix. In
$\overline{DR}$, the mass counterterm $\delta m_A^2$ is chosen in such
a way that cancels exactly the divergences from the self energy. In
addition, the  explicit scale dependence of the self energy is
canceled by the implicit scale dependence of the running mass
parameter $m_A^2(Q)$. In this way, the pole mass $m_A^2$ is finite and
scale independent (up to two-loop  effects).

\section{Low Energy Spectrum}

The two sleptons $\tilde l^{\pm}_L$ and $\tilde l^{\pm}_R$ are the
supersymmetric partners of the left and right handed leptons and, in
general, they can mix. The mass eigenstates are denoted $\tilde
l^{\pm}_1$ and $\tilde l^{\pm}_2$ and the mass matrix is given by
\begin{equation}
{\bold{M_{\tilde l^{\pm}}^2}}=\left[\matrix{
M_L^2+m_l^2-\half(2m_W^2-m_Z^2)c_{2\beta}&
m_l(A_l-\mu t_{\beta})\cr
m_l(A_l-\mu t_{\beta})&
M_R^2+m_l^2-(m_Z^2-m_W^2)c_{2\beta}\cr}\right]\,.
\label{eq:Mslepton}
\end{equation}
In the case of selectrons, mixing can be neglected and the mass
eigenstates are $\tilde e^{\pm}_L$ and $\tilde e^{\pm}_R$. Their
masses, together with the sneutrino mass, are
\begin{eqnarray}
m^2_{\tilde e^{\pm}_L}&=&M_L^2-\half(2m_W^2-m_Z^2)c_{2\beta}\,,
\nonumber \\
m^2_{\tilde e^{\pm}_R}&=&M_R^2-(m_Z^2-m_W^2)c_{2\beta}\,,
\label{eq:selsneum}\\
m^2_{\tilde\nu_e}&=&M_L^2+\half m_Z^2c_{2\beta}\,.\nonumber
\end{eqnarray}
As explained in the previous section, the two soft supersymmetry
breaking terms $M_L^2$ and $M_R^2$ are equal to $m_0$ at the
unification scale (here we are omitting the generation indices). The
solution of the RGE of these two parameters \cite{BBMR}\ can be
approximated by
\begin{eqnarray}
M^2_L(m_Z^2)&\approx&m_0^2+{1\over{16\pi^2}}(3g^2+g'^2)M_{1/2}^2
\ln{{M_X^2}\over{m_Z^2}}\nonumber\\
M_R^2(m_Z^2)&\approx&m_0^2+{1\over{4\pi^2}}g'^2M_{1/2}^2
\ln{{M_X^2}\over{m_Z^2}}\,,
\label{eq:MLMRapp}
\end{eqnarray}
implying that $M_L^2$ is larger than $M_R^2$. In addition, since the
radiative breaking of the electroweak symmetry can be properly
obtained if $\tan\beta>1$, we infer from eq.~(\ref{eq:selsneum}) that
$m_{\tilde e^{\pm}_R}<m_{\tilde e^{\pm}_L}$, and for this reason, we
concentrate on the production of $\tilde e^+_R\tilde e^-_R$ pairs.

Negative searches of charged sleptons
\cite{L3Rep,AlephRep,selchaexp,selexp} impose an experimental lower
bound on their mass given by $m_{\tilde l^{\pm}}>45$ GeV. Because
$\tau^{\pm}$ has a mass much larger than the rest of the leptons, stau
mixing is larger and, consequently, $\tau^{\pm}_1$ is in general the
lightest of the charged sleptons. In this way, the constraint
$m_{\tilde\tau^{\pm}_1}>45$ GeV eliminates part of the parameter
space. On the other hand, a lower bound on the sneutrino mass is
obtained from negative experimental searches
\cite{L3Rep,AlephRep,sneuExp}. If the three sneutrino flavors are
degenerate, the mass has to satisfy $m_{\tilde\nu}>41.8$ GeV, and this
is the case in minimal supergravity because the slepton masses are
equal to $m_0$ at the unification scale and the RGE evolve them almost
equally. This constraint also reduces the allowed parameter space.

The supersymmetric partners of the $W_{\mu}^{\pm}$ gauge bosons and
the charged Higgs bosons mix to form charginos $\chi^{\pm}_{1,2}$,
whose mass matrix is
\begin{equation}
{\bold{M_C}}=\left[\matrix{
 M                    & \sqrt{2}m_Ws_{\beta}\cr
 \sqrt{2}m_Wc_{\beta} & \mu                 \cr
}\right] \; \; .
\label{eq:chamamat}
\end{equation}
This matrix is diagonalized by two unitary matrices $V$ and $U$
chosen such that
\begin{equation}
{\bold U}^*{\bold{M_C}}{\bold V}^{-1}={\bold{M_C^d}}\; ,
\label{eq:UVmat}
\end{equation}
where $M_C^d$ is a diagonal matrix with non-negative entries. The
experimental lower bound on the lightest chargino mass is given by
\cite{L3Rep,AlephRep,selchaexp,chaexp} $m_{\chi^{\pm}_1}>45$ GeV.

In the same way, the supersymmetric partners of the $B_{\mu}$ and
$W^3_{\mu}$ gauge bosons, and the neutral Higgs bosons mix to form
the neutralinos $\chi^0_j$, $j=1,...4$. The mass matrix is
\begin{equation}
{\bold{M_N}}=\left[\matrix{
 M'             & 0              &-m_Zs_Wc_{\beta}& m_Zs_Ws_{\beta}\cr
 0              & M              & m_Zc_Wc_{\beta}&-m_Zc_Ws_{\beta}\cr
-m_Zs_Wc_{\beta}& m_Zc_Wc_{\beta}& 0              &-\mu            \cr
 m_Zs_Ws_{\beta}&-m_Zc_Ws_{\beta}&-\mu            & 0              \cr
}\right] \; \; ,
\label{eq:neutmasmat}
\end{equation}
and can be diagonalized by a matrix $N$ in this basis, or by a
matrix $N'$ in the basis Photino-Zino-Higgsino, according to the
notation of ref.~\cite{GunHaber}. The matrices $N$ and $N'$ can
be complex, and they are chosen in such a way that the eigenvalues
are real and positive. An equivalent way to diagonalize the neutralino
mass matrix is allowing the eigenvalues to be negative: in this case,
the matrix $N$ ($N'$) is replaced by the real matrix $Z$ ($Z'$), and
with the neutralino masses we make the replacement
$m_{\chi^0_i}\longrightarrow\epsilon_i m_{\chi^0_i}$, where
$\epsilon_i$ is the sign of the $i$--th eigenvalue and the neutralino
masses are positive.

Negative experimental searches impose a $\tan\beta$ dependent lower
bound on the lightest neutralino mass \cite{AlephRep,neuexp}. With
the extra assumption of gaugino universality, lower bounds on the
heavier neutralinos (and heavy chargino) can be set \cite{Hidaka}
with the help of gluino searches \cite{gluexp}. Equivalently, we
prefer to impose that the 95\% CL upper bound on the contribution
of new particles to the Z width is $\Delta \Gamma_Z < 23.1$ MeV,
that the branching fraction for the decays $Z\rightarrow
\tilde{\chi}^0_i \tilde{\chi}^0_j$ (where i and j are not both 1)
satisfies $B(Z\rightarrow \tilde{\chi}^0_i \tilde{\chi}^0_j)<10^{-5}$,
and that the invisible width of the $Z$ satisfies
$\Delta\Gamma^{inv}_Z<8.4$ MeV \cite{L3neu}. A light gluino (with a
mass of the order of a few GeV) has not been ruled out by experiments
\cite{LigGluExp}, and gaugino universality implies a lightest
neutralino with a mass smaller than 1 GeV. Nevertheless, in
supergravity models it has been proven that a light gluino is
incompatible with the radiatively broken electroweak symmetry
\cite{MarcoGlu} (see also \cite{OtherGlu}), and we do not consider
this case here.

The Higgs sector of the MSSM consists of five physical states, two
neutral CP-even Higgs bosons $h$ and $H$, one neutral CP-odd Higgs
boson $A$, and a pair of charged Higgs bosons $H^{\pm}$. In an
on-shell scheme, where the CP-odd Higgs mass $m_A$ is defined as the
pole of the propagator and calculated with eq.~(\ref{eq:mApolerun}),
and where $\tan\beta$ is defined through the $A\tau^+\tau^-$ vertex
\cite{diazi}, we compute all the one-loop renormalized Higgs masses.
In the case of the charged Higgs mass $m_{H^{\pm}}$, we include
\cite{diazhaberi} exact one-loop contributions from the top-bottom
quark-squark sector and leading logarithms from lighter generations of
quarks and squarks, three generations of leptons and sleptons, Higgs
and gauge bosons, charginos and neutralinos \cite{ChargedH,chaneu}. We
find that radiative corrections to the charged Higgs mass are small
for the values of $\tan\beta$ we are considering.

In the case of neutral Higgs bosons, radiative corrections are larger
since the leading terms are proportional to $m_t^4$
\cite{chaneu,neutral}. Here we include full one-loop contributions
from top-bottom quarks and squarks and leading logarithms from the
rest of the particles. Two-loop are important \cite{twoloop,RalfHo},
and we include the leading contributions valid at any value of
$\tan\beta$ \cite{RalfHo}. We renormalize the $2\times2$ inverse
propagator matrix of the CP-even neutral Higgs sector using a momentum
dependent mixing angle $\alpha(p^2)$ \cite{marcodpf}, which allows us
to calculate the renormalized value of the parameter
$\sin(\beta-\alpha)$ at the two physical scales $p^2=m_h^2$ and
$p^2=m_H^2$. This parameter is important because it is the MSSM
coupling of $h$ to two $Z$--bosons relative to the $H_{SM}ZZ$
coupling, and in the decoupling limit \cite{HowieDec} where
$\sin(\beta-\alpha)\rightarrow1$, it will be very difficult to
distinguish the MSSM Higgs $h$ from the SM Higgs $H_{SM}$ without
other supersymmetric signal\footnote{Nevertheless, the SM with no new
physics below $\sim10^{10}$ GeV and the MSSM with $M_{SUSY}\lsim1$ GeV
can be distinguished with the measurement of the Higgs mass
\cite{MSSMorSM}.}.

\section{Selectron Pair Production and Decay}

In this section we display the relevant formulas for the total cross
sections of selectron pair production in electron-positron
annihilation and their subsequent decay. In the following, the index
$i$ labels the two different selectrons $\tilde e_i$ with $i=L,R$,
whose masses are denoted by $m_i^2\equiv m_{\tilde e_i}^2$. Similarly,
the index $j$ (as well as $k$) labels the four different neutralinos
$\chi^0_j$ with $j=1,4$, whose masses are $m_j^2\equiv
m_{\chi^0_j}^2$. We start with the production of two selectrons of the
same type. In this case, there are contributions from intermediate
$\gamma$ and $Z$ bosons in the s-channel, and from neutralinos in the
t-channel, as indicated by Figs.~1(a) and 1(b). The two total cross
sections can be calculated with the formula:
\begin{eqnarray}
\sigma(e^+e^-\longrightarrow\tilde e_i^+\tilde e_i^-)&=&
{{(1-4m_i^2/s)^{3/2}}\over{24\pi s}}\Bigg[
{\textstyle{1\over 2}}e^4+
{{g^2a_i^2}\over{16c_W^2}}(1-4s_W^2+8s_W^4)
{{s^2}\over{P_Z(s)}}\nonumber\\
&+&{{e^2ga_i}\over{2c_W}}({\textstyle{1\over 2}}-2s_W^2)
{{s(s-m_Z^2)}\over{P_Z(s)}}\Bigg]
-{1\over{4\pi s}}\sum_{j=1}^4\sum_{k=1}^4|\lambda_{ij}|^2
|\lambda_{ik}|^2 h^{ijk}\nonumber\\
&+&{1\over{8\pi s}}\sum_{j=1}^4 |\lambda_{ij}|^2 f^{ij}
\left[e^2+a_i^2{{s(s-m_Z^2)}\over{P_Z(s)}}\right]
\label{eq:sigmaii}
\end{eqnarray}
where $s$ is the center of mass energy. In the first line of this
formula, the term proportional to $e^4$ corresponds to the photon
contribution, and the term proportional to $g^2$ corresponds to the
$Z$ contribution, where we have defined the constants $a_i$ by
$a_L=g(\half-s_W^2)/c_W$ and $a_R=-gs_W^2/c_W$, and the denominator
$P_Z(s)$ comes from the $Z$ propagator:
$P_Z(s)=(s-m_Z^2)^2+\Gamma_Z^2m_Z^2$. In the second line, the term
proportional to $e^2g$ is the $\gamma-Z$ interference, and the term
with the double sum corresponds to the neutralino sector
($\chi^0$--$\chi^0$ term). Here, $\lambda_{ij}$ are the coeficients
in the $e^+\chi^0_j\tilde e^-_i$ vertex given by
\begin{equation}
\lambda_{Lj}={1\over{\sqrt{2}}}\left[eN'_{j1}+a_L
N'_{j2}\right],\qquad
\lambda_{Rj}=-{1\over{\sqrt{2}}}\left[e{N'}_{j1}^*+a_R
{N'}_{j2}^*\right]
\label{eq:lambdas}
\end{equation}
and the matrix $N'$ is defined below eq.~(\ref{eq:neutmasmat}).

In the $\chi^0$--$\chi^0$ term we also introduce the function
$h^{ijk}$. The definition of this function depends on whether we are
considering the interference between two different neutralinos or
the amplitude squared of one of them. The expression for $h^{ijk}$ is:
\begin{equation}
h^{ijk}=\left\{\begin{array}{ll}
-\half s\frac{f^{ij}-f^{ik}}{m_j^2-m_k^2}
& \quad if\quad m_j^2\ne m_k^2\\
\,\,\,\, g^{ij} & \quad if \quad m_j^2=m_k^2\\
\end{array}\right.
\label{eq:hijk}
\end{equation}
where the two functions $f^{ij}$ and $g^{ij}$ are given by
\begin{equation}
f^{ij}=\sqrt{1-4m_i^2/s}\left(-1+2{{\Delta m_{ij}^2}\over s}\right)
+2\left[{{m_j^2}\over s}+{{(\Delta m_{ij}^2)^2}\over{s^2}}\right]
\ln\left[{{\Delta m_{ij}^2-\half sy_+}\over{\Delta m_{ij}^2
-\half sy_-}}\right]
\label{eq:fij}
\end{equation}
and
\begin{equation}
g^{ij}=2\sqrt{1-4m_i^2/s}+\left(-1+2{{\Delta m_{ij}^2}\over s}\right)
\ln\left[{{\Delta m_{ij}^2-\half sy_+}\over{\Delta m_{ij}^2-\half
sy_-}}\right] \;\; .
\label{eq:gij}
\end{equation}
In this case, we have defined $\Delta m_{ij}^2=m_i^2-m_j^2$ and
$y_{\pm}=1\pm\sqrt{1-4m_i^2/s}$. Finally, the third line in
eq.~(\ref{eq:sigmaii}) has the $\gamma$--$\chi^0$ interference,
proportional to $e^2$, and the $Z$--$\chi^0$ interference,
proportional to $a_i^2$. In these terms, the function $f^{ij}$ is
the one defined in eq.~(\ref{eq:fij}).

Now we turn to the production of two selectrons of a different
type. The total cross sections receive contributions only from
neutralinos in the t-channel [Fig.~1(b)], and the formulas are
simply:
\begin{equation}
\sigma(e^+e^-\longrightarrow\tilde e_L^+\tilde e_R^-)=
\sigma(e^+e^-\longrightarrow\tilde e_R^+\tilde e_L^-)=
{1\over{4\pi s}}\sum_{j=1}^4\sum_{k=1}^4\lambda_{Lj}\lambda_{Rj}
\lambda_{Lk}^*\lambda_{Rk}^*H^{jk}
\label{eq:sigmaLR}
\end{equation}
where the couplings $\lambda_{ij}$ are defined in
eq.~(\ref{eq:lambdas}), and the function $H^{jk}$ is given by:
\begin{equation}
H^{jk}=\left\{\begin{array}{ll}
-{{m_jm_k}\over{m_j^2-m_k^2}}\left(\ln\left[
{{\Delta{\overline m}_j^2-s\bar y_+/2}\over{\Delta{\overline m}_j^2-
s\bar y_-/2}}\right]-
\ln\left[{{\Delta{\overline m}_k^2-s\bar y_+/2}\over{
\Delta{\overline m}_k^2-s\bar y_-/2}}\right]\right)
& \quad if\quad m_j^2\ne m_k^2\\ \\
\,\,\,\,\lambda_{ad}^{1/2}{{m_j^2s}\over{
(\Delta{\overline m}_j^2-s\bar y_+/2)
(\Delta{\overline m}_j^2-s\bar y_-/2))}}
& \quad if \quad m_j^2=m_k^2\\
\end{array}\right.
\label{eq:Hjk}
\end{equation}
This time we define
$\Delta{\overline m}_j^2=\half(m_L^2+m_R^2)-m_j^2$ and
$\bar y_{\pm}=1\pm\lambda_{ad}^{1/2}$. The function $\lambda$ is
well known, $\lambda(a,b,c)=a^2+b^2+c^2-2ab-2ac-2bc$ and the subscript
``$ad$'' is used to remind the reader that we have chosen to use
adimensional arguments: $\lambda_{ad}=\lambda(1,m_L^2/s,m_R^2/s)$.

The main decay modes of a selectron are neutralinos
$\tilde e^{\pm}_i\rightarrow e^{\pm}\chi^0_j$, charginos
$\tilde e^{\pm}_i\rightarrow \nu_e\chi^{\pm}_k$, and sneutrinos
$\tilde e^{\pm}_i\rightarrow W^{\pm}\tilde\nu_e$, where $i=L,R$,
$j=1,...4$, and $k=1,2$. The corresponding decay rates are:
\begin{eqnarray}
\Gamma(\tilde e^-_i\longrightarrow e^-\chi^0_j)&=&{1\over{4\pi}}
|\lambda_{ij}|^2m_i\left(1-{{m_j^2}\over{m_i^2}}\right)^2\nonumber\\
\Gamma(\tilde e^-_L\longrightarrow \nu_e\chi^-_k)&=&{g^2\over{16\pi}}
|U_{k1}|^2m_L\left(1-{{m_k^2}\over{m_L^2}}\right)^2
\label{eq:decayneut}\\
\Gamma(\tilde e^-_L\longrightarrow W^-\tilde\nu_e)&=&
{{g^2}\over{32\pi}}{{m_L^3}\over{m_W^2}}
\lambda^{3/2}(1,m_{\tilde\nu_e}^2/m_L^2,m_W^2/m_L^2)
\nonumber
\end{eqnarray}
where $\lambda_{ij}$ is defined in eq.~(\ref{eq:lambdas}), the matrix
$U$ is defined in eq.~(\ref{eq:UVmat}), and the function
$\lambda(a,b,c)$ is defined below eq.~(\ref{eq:Hjk}).

\section{Results}

Once we have introduced the theoretical framework in which we want to
work, we can proceed to present some results; as was mentioned at the
end of section 2, our free high--energy parameters, once we take into
account the constraints imposed by a correct electroweak breaking, are
$m_0$, $M_{1/2}$, $\tan \beta$ and sign($\mu$). Let's restrict
ourselves, for the moment, to the $\mu<0$ case: this leaves us with
only three parameters, $M_{1/2}$, $m_0$ and $\tan \beta$. For the
purposes of our calculation (that is, to find regions with a
non--negligible selectron pair production cross--section) we must
choose those values of $M_{1/2}$, $m_0$ which give rise to the lowest
possible values for the selectron masses. Let's stress here that, in a
SUGRA model, all the spectrum is correlated, so that every choice of
$M_{1/2}$, $m_0$ and $\tan \beta$ determine the values the masses of
{\em all} the susy particles; so we have to keep in mind their
experimental constraints \cite{PDG} simultaneously.

The regions we have explored can be labeled by their $\tan\beta$
value; once this is specified we have found more intuitive to classify
the spectrum by their value for the common gaugino mass, $M_{1/2}$.
This is due to the fact that, given a value of $\tan\beta$,
the neutralino and chargino mass matrices
are determined by the values of $M_{1/2}$ and $\mu$. In these
minimal SUGRA models we know that the value of $\mu$ is given by the
requirement of a correct elecroweak breaking (i.e. a correct $M_Z$),
and in fact we have observed that, despite the value of $m_0$ that we
are considering, we need $\mu$ to be always larger than $M_{1/2}$.
What this is telling us is that the lightest neutralino and chargino
are going to be mainly gaugino, so that the lowest experimental bounds
on these particles (mainly on neutralinos through the experimental
restrictions imposed to the decay of the $Z$ in visible neutralinos,
see sect.~3) set a lower bound on the value of $M_{1/2}$ that we are
able to take. And this will be between 70 and 100 GeV. In conclusion,
given a choice for $\tan\beta$, all the points in our parameter space
with a common value for the gaugino mass share the same value for the
lightest neutralino and chargino ones, and also for that of the gluino
(note that the latter is even almost independent of the value of $\tan
\beta$, as can be seen in Table~1). In our case, and for reasons that
will become clear soon, we have explored a region of values for
$M_{1/2}$ up to 150 GeV.
\begin{center}
\begin{tabular}{|c|c|c|c|c|} \hline
$M_{1/2}$ & $\tan\beta$ & $m_{\chi^0_1}$ &
$m_{\chi^{\pm}_1}$ & $m_{\tilde g}$ \\ \hline
 70 &  2 & 32 &  77 & 224 \\ \hline
 80 &  2 & 37 &  83 & 255 \\ \hline
 90 &  2 & 41 &  90 & 286 \\ \hline
100 &  2 & 45 &  97 & 317 \\ \hline
150 &  2 & 65 & 135 & 480 \\ \hline
 90 & 10 & 35 &  59 & 286 \\ \hline
100 & 10 & 40 &  69 & 318 \\ \hline
120 & 10 & 48 &  87 & 381 \\ \hline
150 & 10 & 61 & 114 & 478 \\ \hline
100 & 25 & 37 &  62 & 318 \\ \hline
120 & 25 & 46 &  81 & 382 \\ \hline
\end{tabular}
\vskip .5cm
{\footnotesize
Table 1. Lightest neutralino, lightest chargino and gluino masses
(in GeV) for different choices of $M_{1/2}$ and $\tan \beta$, and
$\mu<0$. These masses are independent of the value of $m_0$, and
because of this, any of them can be used to label the curves we show
in Figs.~2--8.}
\end{center}
But we still cannot say much about scalar masses, in particular about
selectron masses, as we have not specified any value at all for $m_0$.
This will be given, in its lowest bound, by the experimental bounds on
the slepton masses, the lightest among the scalars. For low values of
$\tan\beta$ we have a very small mixing in the stau mass matrix,
therefore the lightest scalar is the sneutrino (being all selectrons,
smuons and staus almost degenerate in mass). Let's note that, as
$M_{1/2}$ increases, this lower bound on $m_0$ decreases, due to the
presence of gaugino masses in the RGEs of scalar masses \cite{BBMR}
as indicated by eq.~(\ref{eq:MLMRapp}); in fact for $\tan\beta=2$ and
$M_{1/2}\simgt100$ GeV we find that it practically disappears (i.e.
any choice for $m_0$ gives a spectrum compatible with the experimental
bounds).

The behaviour of the different low--energy masses is presented in
Figs.~2--6, where the curves are labeled by the corresponding lightest
neutralino mass (or, equivalently, by the value of $M_{1/2}$, as can be
seen in Table 1). In Figs.~2 and 3 we have plotted the sneutrino (the
three sneutrino species are practically degenerate) and lightest stau
masses respectively, and we can clearly appreciate how the experimental
lower bounds on their masses restrict the allowed parameter space.
In cases (a) $\tan\beta=2$ and (b) $\tan\beta=10$, restrictions coming
from the sneutrino mass dominate as we see in Fig.~2, where some of the
curves are truncated by the condition $m_{\tilde\nu}>41.8$ GeV.
We can also see that the range of right--handed selectron mass
(plotted in the x--axis), which corresponds to a particular choice of
$m_{\chi^0_1}$, increases with this mass until it reaches a maximum.
The reason for this is the decrease of the lower bound on $m_0$
mentioned before: since $m_{\tilde\nu}$ receives contributions from
$M_{1/2}$ as well as from $m_0$, higher values of $M_{1/2}$ allow
lower values of $m_0$. From eqs.~(\ref{eq:selsneum}) and
(\ref{eq:MLMRapp}) we get:
\begin{equation}
m_{\tilde e_R^{\pm}}^2\approx m_{\tilde\nu}^2-{3\over{16\pi^2}}
(g^2-g'^2)M_{1/2}^2\ln{{M_X^2}\over{m_Z^2}}-\half(3m_Z^2-2m_W^2)
c_{2\beta},
\label{eq:relmsemsn}
\end{equation}
and from this equation we see that for a constant value of
$\tan\beta$, and for $m_{\tilde\nu}=41.8$ GeV, the minimum selectron
mass decreases with an increasing $m_{\chi^0_1}$. After its maximum,
the range of $m_{\tilde e_R^{\pm}}$ becomes narrower because
the value of $M_{1/2}$ is big enough to be unable to produce
light selectrons anymore. This is the reason to limit the plots
to $M_{1/2} \leq 150$ GeV.

As the value of $\tan\beta$ increases, the stau mixing also does, and
experimental restrictions coming from the stau mass start to dominate;
in fact, it can be the case that the stau is lighter than the lightest
neutralino, $\chi_1^0$, becoming thus the lightest supersymmetric
particle (LSP). This situation is cosmologically disfavoured
\cite{LSP}, so it provides us with a new constraint on our parameter
space, which is best appreciated in Fig.~3. For large $\tan \beta$ the
lightest stau becomes also the lightest scalar, and we can see from
Fig.~3(c) that one of the curves is truncated by the condition
$m_{\tilde\tau^{\pm}_1}>45$ GeV. However, as $M_{1/2}$ increases the
mass of the lightest neutralino also does, and for $m_{\chi_1^0}>45$
GeV, the condition that the LSP has to be electrically neutral imposes
$m_{\tilde\tau^{\pm}_1}>m_{\chi_1^0}$. In fact, this latter
requirement rules out any spectra with $M_{1/2}=150$ GeV and
$\tan\beta=25$. Also, in this large $\tan\beta$ regime it is no longer
possible to reach values for $m_0$ as low as in the former cases
(where we had small or zero stau mixing). For example, for
$\tan\beta=25$ and $M_{1/2} \geq 100$ GeV, $m_0$ must be always bigger
than $55$ GeV\footnote{
This can be easily seen if we rewrite the
condition $m_{\tilde\tau^{\pm}_1}^2>m_{\chi_1^0}^2$ in terms of the
corresponding RGEs for large $\tan \beta$, and taking into account
that, in our model, the lightest neutralino is mainly gaugino. Then the
relationship between the different parameters is given by:
$m_0^2+0.23 M_{1/2}^2 > 1.75 |\mu(m_Z)| \tan \beta$.}.
The absence of low
values for $m_0$ sets a minimum in the range of selectron masses that
increases uniformly with increasing $m_{\chi^0_1}$.

In Fig.~4 we plot the lightest top and bottom squark masses. Squarks
are heavier than sleptons mainly because the RGE of soft squark masses
receive contributions from the strong coupling constant. In general
bottom squarks are heavier that top squarks, and the reasons are that:
i) the sbottom soft mass is larger than the stop one because Yukawa
couplings contribute negatively to their RGEs, and ii) the
off-diagonal elements of the stop mass matrix (which are proportional
to the corresponding Yukawa coupling and $\tan \beta$) are much
bigger. The exception occurs in case (a) with light neutralino, where
$m_{\tilde b_1}$ is smaller than $m_{\tilde t_1}$, the reason being
that the soft squark masses are comparable to $m_t$, so that the
diagonal elements of the stops are much bigger than those of the
sbottoms. In addition, the small value of $\tan\beta$ makes the
off--diagonal term less important (see the squark mass matrices in the
appendix).

The Higgs spectrum is presented in Figs.~5 and 6. The masses of the
heavy Higgs bosons, given by the CP-odd Higgs $A$, the charged Higgs
$H^{\pm}$, and the heavy CP-even Higgs $H$, are plotted in Fig.~5.
In the region of parameter space we are considering here, radiative
corrections to $m_{H^{\pm}}$ and $m_H$ are small, typically not larger
than a few GeV. Consequently, the charged Higgs and heavy CP-even
Higgs pole masses are, to a good approximation, determined by the
values of the pole mass $m_A$ and $\tan\beta$. Similarly, the
difference between the pole mass $m_A$ and the running mass
$m_A(\widehat Q)$ is small. On the other hand,
the running CP-odd Higgs mass $m_A(\widehat Q)$ is determined by the
correct radiatively broken electroweak symmetry ($\tan\beta$ is
an input). We find that for $\tan\beta=2$, the CP-odd Higgs is heavier
in comparison with higher values of $\tan\beta$, and that increasing
values of $M_{1/2}$ produce higher values of $m_A$.

The lightest Higgs $h$ is massless at tree level if $\tan\beta=1$, and
its mass increases until it saturates as $\tan\beta\rightarrow\infty$.
This effect can be appreciated as we compare the three cases
$\tan\beta=$2, 10, and 25 in Fig.~6. Two different effects influence
the dependence of the Higgs mass $m_h$ on the lightest neutralino mass.
First, if $m_{\chi^0_1}$ increases $m_A$ also does, producing higher
values of $m_h$. Second, larger values of squark masses are obtained
if $m_{\chi^0_1}$ increases, producing larger contributions to $m_h$
from radiative corrections. This latest effect can be seen in
Fig.~6(c), because at large $\tan\beta$, the lightest Higgs mass $m_h$
becomes independent of $m_A$ at tree level, and equal to $m_Z$. The
difference is due to radiative corrections. From this figure it is
obvious the importance of the inclusion of radiative corrections in
the calculation of $m_h$.

A SM Higgs boson can be detected at LEPII if its mass is smaller than
about 105 GeV \cite{SMhiggs} (brehmsstrahlung of a Higgs by a $Z$
gauge boson). Since the lightest Higgs mass $m_h$ in Fig.~6 is always
lighter than about 110 GeV, if this scenario is correct, it is likely
to be observed at LEPII. Production cross sections and decay rates of
this supersymmetric Higgs $h$ are close to the ones of the SM Higgs,
since the parameter $\sin(\beta-\alpha)$ is always close to one,
making difficult to distinguish between the two models
\cite{HowieDec,MSSMorSM}. Nevertheless, for values of
$m_{\chi^0_1}\lsim 40$ GeV and $\tan\beta=10$ or 25, this parameter
can be as small as $\sin(\beta-\alpha)\gsim 0.95$, which means that
the ratio between the production cross sections of the two models can
be as low as $\sigma_{MSSM}/\sigma_{SM}\gsim 0.9$ and discrimination
may be possible. Production of heavy CP-even Higgs bosons at LEPII
will be difficult. In the cases described above, $m_H$ is about
120--130 GeV and $|\cos(\beta-\alpha)|\lsim 0.3$, which means that
its production cross section is about 10\% of the SM one.

Once we have calculated both the spectrum and the total cross section,
we can present our predictions. Those are shown in Figs.~7(a)--7(c)
where we have plotted the total cross section
$\sigma(e^+e^-\longrightarrow\tilde e_R^+\tilde e_R^-)$ (the only
relevant one) versus the right-handed selectron mass for a
center--of--mass energy of 200 GeV and several choices of the lightest
neutralino mass (see caption for details). As was mentioned before, low
values for $\tan\beta$ produce smaller values for $m_{\tilde
e_R^{\pm}}$, and therefore a higher total cross section. In general we
see that, for every $\tan \beta$ value we have examined, the different
curves (labeled by their corresponding $m_{\chi_1^0}$ value) tend to
be very close to each other making potentially very difficult to
identify a particular minimal SUGRA spectrum as the one corresponding
to a certain value of both $m_{\tilde e_R^{\pm}}$ and
$\sigma(e^+e^-\longrightarrow\tilde e_R^+\tilde e_R^-)$. But luckily
enough we are provided with another tool to tell them apart.

As has been stressed in recent years, the $b \rightarrow s, \gamma$
decay is becoming a very powerful test of physics beyond the Standard
Model \cite{BBMR,BSG,BdeCyC}, in particular since the measurement of
its branching ratio (BR) by the CLEO collaboration. As a FCNC
process, $b \rightarrow s,\gamma$ is forbidden at tree level, so
1--loop diagrams become the dominant ones. In our case the presence
of susy particles provides us with an extra contribution to this BR
to be added to the usual SM one (given mainly by a top quark and a
$W^{-}$ boson running in the loop); therefore the actual experimental
bounds $1 \times 10^{-4} \leq {\rm BR}(b \rightarrow s, \gamma) \leq
4.2 \times 10^{-4}$ \cite{CLEO} put strong constraints on the
supersymmetric spectrum (for a detailed analysis see
refs.~\cite{BBMR,BSG,BdeCyC}), and we have used this fact to discard
some of the spectra we have shown in the previous plots. As we can see
in Fig.~8, where the predicted BR is plotted for each spectrum we have
considered versus the corresponding selectron mass, the bigger $\tan
\beta$ is, the less compatible with the experiment the spectra are.
While in Fig.~8(a) we see that the lightest neutralino mass curve
is very close to the SM prediction, as $\tan \beta$ increases the
corresponding line in Fig.~8(b) comes closer to the lower CLEO bound
and finally, in 8(c), goes even below it, becoming totally discarded
(that being the reason why it has been represented by a dashed--dotted
line in Fig.~7).

Pretty much the same happens with the highest neutralino mass line:
for increasing $\tan \beta$ it becomes closer and even above the upper
CLEO bound. Nevertheless, we are being conservative in our predictions
and, considering the estimated 25 $\%$ error in the theoretical
calculation for the BR($b \rightarrow s, \gamma$), we have kept the
highest neutralino mass line for $\tan\beta=10$ as compatible with the
bound\footnote{Remember that the equivalent case for $\tan \beta=25$
was discarded for giving a charged (lightest stau) LSP.} (and
therefore represented as a solid line in Figs.~7(b) and 8(b)). In any
case it is noticeable how an improvement of the experimental
measurement by an order of magnitude would discard most of the
parameter space shown in our plots. Note also how the closest
prediction to that of the SM corresponds to a heavier spectrum as
$\tan\beta$ increases: for $\tan\beta=2$ we have $m_{\chi^0_1}=32$
GeV, while for $\tan\beta=10$ we get $m_{\chi^0_1}=40$ GeV and for
$\tan\beta=25$, $m_{\chi^0_1}=46$ GeV. That, combined with the results
shown in Fig.~7, tells us that if minimal SUGRA is the theory beyond
the SM, a low $\tan\beta$ value is more likely to be discovered at
LEPII, since it predicts a much lighter spectrum and a sizable cross
section for selectron production than higher values of $\tan\beta$;
alternatively the reverse statement does not have to be true, that is
the absence of measurements at LEPII would not imply a particular
range of values of $\tan \beta$ as the preferred one.

Once selectrons are produced we can examine the decay channels in this
Minimal SUGRA model. As we explained in the last section, there are
three main possible ways for this to happen. However, taking into account
that we are focussing in right handed selectrons (the lightest and,
therefore the first ones to be produced), two of the couplings
($\tilde e^-_R W^+ \tilde \nu_e$ and $\tilde
e^-_R \nu_e \chi^+_k$) are zero and we are left with
neutralinos as the only decaying products. In the scenario we consider
here, $\chi^0_3$ and $\chi^0_4$ are too heavy, and these decays are
kinematically forbidden. In Fig.~9 we plot the branching ratios of the
remaining two decay modes, namely $\tilde e^{\pm}_R\rightarrow
e^{\pm}\chi^0_j$, with $j=1,2$. We do not consider cascade decays
\cite{Cuypers}. The branching ratios are plotted as a function of the
selectron mass, in the case $\mu<0$, and for different values of
$\tan\beta$ and $M_{1/2}$. Note that the decay rate of a right
selectron to the lightest neutralino becomes dominant at low values of
$\tan\beta$ (case 1), where the difference between $m_{\chi_1^0}$ and
$m_{\chi_2^0}$ is more pronounced (this can be checked by looking at
Table~1, and taking into account that $m_{\chi_2^0}$ is practically
identical to $m_{\chi_1^{\pm}}$). On the other hand, the decay into
the second lightest neutralino is important at high values of
$\tan\beta$ and low universal gaugino mass $M_{1/2}$ (cases 2 and 4)
for which the magnitudes of the two neutralino masses are closer.

Let's turn now to $\mu>0$. In this case we have observed a much bigger
suppression of the allowed parameter space, as becomes clear from
Table~2:
\begin{center}
\begin{tabular}{|c|c|c|c|c|}\hline
$M_{1/2}$ & $\tan \beta$ & $m_{\chi_1^0}$ & $m_{\chi_1^{\pm}}$ &
$m_{\tilde g}$ \\ \hline
120 & 2  & 34 & 64  & 382 \\ \hline
150 & 2  & 51 & 94  & 481 \\ \hline
100 & 10 & 30 & 52  & 318 \\ \hline
120 & 10 & 41 & 71  & 381 \\ \hline
150 & 10 & 56 & 100 & 479 \\ \hline
100 & 25 & 33 & 56  & 318 \\ \hline
120 & 25 & 43 & 74  & 382 \\ \hline
\end{tabular}
\vskip .5cm
\footnotesize{
Table 2. Lightest neutralino, lightest chargino and gluino masses (in
GeV) for different choices of $M_{1/2}$ and $\tan \beta$, and $\mu>0$.}
\end{center}
Compared to the $\mu<0$ case, to produce a spectrum (in particular
charginos and neutralinos) which is compatible with all the
experimental bounds on susy masses and precision measurements on the
$Z$ peak, much bigger values of $M_{1/2}$ are required. In addition, the
minimum value of $M_{1/2}$ that fulfills these requirements decreases
with increasing $\tan \beta$ as opposed to what happened with $\mu<0$.
Concerning the range of variation of $m_0$, it is similar to what we
found before: no bound for $\tan \beta=2$, while for $\tan \beta=10$ a
too light sneutrino imposes $m_0 \geq 30$ GeV for $M_{1/2}=100$ GeV
(and no restrictions for $M_{1/2}=120,150$ GeV), and for
$\tan\beta=25$ a too light stau implies $m_0 \geq 55$ GeV for any
value of $M_{1/2}$. Furthermore, any spectra with $\tan\beta=25$ and
$M_{1/2}=150$ GeV is ruled out considering that the LSP is electrically
neutral, as it happened with $\mu<0$. In general we have found that,
for a fixed $M_{1/2}$, the spectrum is insensitive to sign$(\mu)$
when $\tan\beta$ is large, while it becomes lighter for $\mu>0$ as
$\tan\beta$ decreases.

Because values of $M_{1/2}$ as small as in the previous case are not
allowed, the CP-odd Higgs mass necessary to obtain a correct
electroweak symmetry breaking is larger when $\mu>0$ and,
consequently, the charged Higgs $H^{\pm}$ and the CP-even Higgs $H$
are heavier. At the same time, the light CP-even Higgs $h$ is heavier
because $m_A$ is larger (increasing the tree level mass) and because
squarks are heavier (increasing the radiative corrections). In
addition, when $\mu>0$ the Higgs $h$ behaves more like the SM Higgs
because a larger $m_A$ produces a value of $\sin(\beta-\alpha)$ closer
to unity. Nevertheless, still there are good chances to detect this
Higgs boson since its mass remains below 110 GeV.

The total cross section versus the right selectron mass for this case
and $\tan \beta=2,10,25$ is presented in Fig.~10(a). As we can see,
the most relevant cross sections are already ruled out by considering
the restrictions coming from the $b \rightarrow s, \gamma$ decay,
Fig.~10(b). As was shown in ref.~\cite{BdeCyC}
\footnote
{We advise the reader that this reference uses the opposite convention
on the sign of $\mu$, compared with the one we use here.},
for $\mu>0$ there are no low--energy windows, that is, regions of the
parameter space which give a very light spectrum and still are
compatible with the experimental measurement for the BR$(b \rightarrow
s,\gamma)$. Therefore the behaviour of the curves is much more uniform
than in the previous case and the bigger $M_{1/2}$ is, the closer its
associated BR is to the SM prediction. In this particular case we see
that $M_{1/2}=120$ GeV for $\tan \beta=2,10$, and $M_{1/2}=140$ GeV
for $\tan \beta=25$ are the lowest possible values for which their BR
start to be compatible with the experimental bound. However an
improvement of this measurement would soon discard most of these
scenarios with $\mu>0$, having perhaps those with $\tan \beta=2$ as
the only surviving ones. In any case let's stress the fact that, in
the region of the parameter space so far analyzed, $\mu<0$ gives a
much lighter spectrum and therefore a bigger cross section for the
production of selectrons.

If right selectrons are discovered at LEPII, measurements of the total
cross section $\sigma(e^+e^- \longrightarrow\tilde{e}^+_R
\tilde{e}^-_R)$, the selectron mass $m_{\tilde e^{\pm}_R}$, and the
mass of the lightest neutralino $m_{\chi^0_1}$ will be available with
some degree of experimental error. By looking at Fig.~11 we can see
how these measurements can be used to test the model. In this figure
we plot the total cross section as a function of the neutralino mass
for constant values of the selectron mass $m_{\tilde e^{\pm}_R}=78$
and 86 GeV, and both signs of $\mu$. In each case, curves of constant
$\tan\beta$ are shown. We appreciate that the dependence of the cross
section both on $\tan\beta$ and sign($\mu$) is weak, therefore, in a
first stage where experimental errors may be high, we will not be able
to predict the value of these important parameters from these
measurements alone\footnote{
The observation of the lightest Higgs, for
example, can give us information on $\tan\beta$, since $m_h$ depends
strongly on this parameter, as we can see from Fig.~6.}.
Nevertheless, it will be possible to test the model, because the
curves corresponding to one value of $m_{\tilde e^{\pm}_R}$ are close
together and reasonably apart from the group of curves corresponding
to the other value of $m_{\tilde e^{\pm}_R}$. For this to happen,
there is a maximum tolerable experimental error on the selectron mass
which is obvious from the figure. In the next experimental stage,
where precision measurements are available, it will be possible to
predict both the values of $\tan\beta$ and sign($\mu$).

\section{Conclusions}

Despite of the multi-dimensional free parameter space that Minimal
Supergravity predicts, many phenomenological restrictions can be
used altogether to reduce the possible values of the soft terms that
define the spectrum. Apart from the direct experimental limits on susy
particles and Z--pole physics, we have also imposed constraints
coming from a prediction of these spectra for the BR of $b \rightarrow
s, \gamma$ compatible with the experimental value, and from the
requirement that the LPS should be neutral.

The discovery of selectrons at LEPII will allow us to test
the remaining Minimal SUGRA scenarios. In a stage where experimental
errors are large, LEP measurements, like the total production cross
section $\sigma(e^+e^- \longrightarrow\tilde{e}^+_R \tilde{e}^-_R)$,
the selectron mass $m_{\tilde e^{\pm}_R}$, and the lightest neutralino
mass $m_{\chi^0_1}$, will validate or rule out the model in its
simplest form. In this scenario, other supersymmetric particles are light
enough to be produced at LEPII. Among them is the lightest CP-even
neutral Higgs boson $h$, whose discovery and measurement of its mass
can give us an important insight into the value of $\tan\beta$ and
sign($\mu$). On the other hand, in a stage where precision
measurements of the observables mentioned above are available,
information on the value of $\tan\beta$ and the sign of $\mu$ can
be directly obtained, and with it, the entire supergravity spectrum can be
predicted.

\vspace{0.25in}

\section*{Acknowledgements}

The work of BdeC was supported by a Spanish MEC Postdoctoral
Fellowship for most of the time while this research was carried out,
and by a PPARC Postdoctoral Fellowship during the final stages of it.

\section*{Appendix}

The exact one-loop contribution to the CP-odd Higgs self energy
from loops involving top and bottom quarks and squarks is
given in this appendix using the following independent
Veltman's functions \cite{Veltman}:

\begin{eqnarray}
A_0(m^2)&=&m^2[\Delta -\ln(m^2/Q^2)+1]\nonumber\\ \\
B_0(p^2;m_1^2,m_2^2)&=&\Delta-\int_0^1 dx
\ln\{[m_2^2 x+m_1^2(1-x)-p^2x(1-x)-i\varepsilon]/Q^2\}\nonumber
\label{eq:a0b0}
\end{eqnarray}
where $Q$ is an arbitrary mass scale,
$\Delta$ is the regulator of dimensional regularization
given by
\begin{equation}
\Delta={2\over{4-n}}+\ln 4\pi-\gamma_E,
\label{eq:delta}
\end{equation}
$n$ is the number of space-time dimensions, and $\gamma_E$ is
the Euler's constant.

We denote the sum of the Feynman diagrams contributing to the
CP-odd Higgs self energy by $-iA_{AA}(p^2)$. In order to
save space we will use the following notation
for the Veltman's functions: $A_0(m_a^2)\equiv A_0^a$ and
$B_0(p^2;m_a^2,m_b^2)\equiv B_0^{pab}$ where $a$
and $b$ are any particle we are considering.

We begin with
the contribution to the CP-odd Higgs self energy
from loops involving top and bottom quarks:
\begin{equation}
\left[A_{AA}(p^2)\right]^{tb}={{N_cg^2m_t^2}\over
{32\pi^2m_W^2t_{\beta}^2}}\left(2A_0^t-p^2B_0^{ptt}
\right)+{{N_cg^2m_b^2t_{\beta}^2}\over
{32\pi^2m_W^2}}\left(2A_0^b-p^2B_0^{pbb}
\right)
\label{eq:Atb}
\end{equation}
with $N_c=3$ being the number of colors. In the same way,
the contribution from top and bottom squarks is
\begin{eqnarray}
\left[A_{AA}(p^2)\right]^{\tilde t\tilde b}&=&
-{{N_c}\over{16\pi^2}}\sum_{i=1}^2\sum_{j=1}^2
(M_{A\tilde t\tilde t}^{ij})^2
B_0^{p\tilde t_i\tilde t_j}+{{N_c}\over{16\pi^2}}
\sum_{i=1}^2(\lambda_{AA\tilde t\tilde t}^{ii})^2
A_0^{\tilde t_i}\nonumber\\
&&-{{N_c}\over{16\pi^2}}\sum_{i=1}^2\sum_{j=1}^2
(M_{A\tilde b\tilde b}^{ij})^2
B_0^{p\tilde b_i\tilde b_j}+{{N_c}\over{16\pi^2}}
\sum_{i=1}^2(\lambda_{AA\tilde b\tilde b}^{ii})^2
A_0^{\tilde b_i}
\label{eq:Astsb}
\end{eqnarray}
The matrices
$M_{A\tilde t\tilde t}$ and $M_{A\tilde b\tilde b}$ correspond
to the numerical factor in the $A\tilde t\tilde t$
and $A\tilde b\tilde b$ Feynman
rules in the basis where the squark
mass matrices are diagonal. They are given by
\begin{equation}
M_{A\tilde t\tilde t}={{gm_t}\over{2m_W}}(\mu+A_t/t_{\beta})
\left[\begin{array}{cc}0&-1\\
                       1&\,\,0
      \end{array}
\right],\quad
M_{A\tilde b\tilde b}={{gm_b}\over{2m_W}}(\mu+A_bt_{\beta})
\left[\begin{array}{cc}0&-1\\
                       1&\,\,0
      \end{array}
\right]
\label{eq:Mattabb}
\end{equation}
In the same way, the matrices $\lambda_{AA\tilde t\tilde t}$
and $\lambda_{AA\tilde b\tilde b}$ are
the numerical factor in the $AA\tilde t\tilde t$
and the $AA\tilde b\tilde b$ vertices in
the physical basis, and they are given by
\begin{equation}
\lambda_{AA\tilde t\tilde t}= R_t\lambda'_{AA\tilde t\tilde t}
R^{-1}_t,\quad\lambda_{AA\tilde b\tilde b}= R_b
\lambda'_{AA\tilde b\tilde b}R^{-1}_b
\label{eq:lAAttAAbb}
\end{equation}
where $\lambda'_{AA\tilde t\tilde t}$ and
$\lambda'_{AA\tilde b\tilde b}$ are the equivalent
matrices but in the $\tilde t_L$--$\tilde t_R$
and $\tilde b_L$--$\tilde b_R$ basis
\begin{eqnarray}
\lambda'_{AA\tilde t\tilde t}&=&\half g^2\left[
\begin{array}{cc}{{1/2-e_ts_W^2}\over{c_W^2}}
c_{2\beta}-{{m_t^2}\over{m_W^2t_{\beta}^2}}&0\\
0&{{e_ts_W^2}\over{c_W^2}}c_{2\beta}-
{{m_t^2}\over{m_W^2t_{\beta}^2}}
\end{array}\right]\nonumber\\ \\
\lambda'_{AA\tilde b\tilde b}&=&\half g^2\left[
\begin{array}{cc}-{{1/2+e_bs_W^2}\over{c_W^2}}
c_{2\beta}-{{m_b^2t_{\beta}^2}\over{m_W^2}}&0\\
0&{{e_bs_W^2}\over{c_W^2}}c_{2\beta}-
{{m_b^2t_{\beta}^2}\over{m_W^2}}
\end{array}\right],\nonumber
\label{eq:lpAAttbb}
\end{eqnarray}
and we use the notation $e_t=2/3$, $e_b=-1/3$,
$t_{\beta}=\tan(\beta)$, and $c_{2\beta}=\cos(2\beta)$.
$R_t$ and $R_b$ are rotation matrices defined by
\begin{equation}
R_t=\left[\begin{array}{cc}\cos(\alpha_t)&\sin(\alpha_t)\\
                          -\sin(\alpha_t)&\cos(\alpha_t)
          \end{array}\right],\quad
R_b=\left[\begin{array}{cc}\cos(\alpha_b)&\sin(\alpha_b)\\
                          -\sin(\alpha_b)&\cos(\alpha_b)
          \end{array}\right],
\label{eq:RtRb}
\end{equation}
which diagonalize the squark mass matrices
\begin{equation}
R_tM_{\tilde t}^2R_t^{-1}=\left(\begin{array}{cc}
                          m_{\tilde t_1}^2&0\\
                          0&m_{\tilde t_2^2}
                                \end{array}\right),\quad
R_bM_{\tilde b}^2R_b^{-1}=\left(\begin{array}{cc}
                          m_{\tilde b_1}^2&0\\
                          0&m_{\tilde b_2^2}
                                \end{array}\right),
\label{eq:Msqdiag}
\end{equation}
given by
\begin{eqnarray}
M^2_{\tilde t}&=&\left[\begin{array}{cc}
    M_Q^2+m_t^2+\sixth(4m_W^2-m_Z^2)c_{2\beta}&
    m_t(A_t-\mu/t_{\beta})\\
    m_t(A_t-\mu/t_{\beta})&
    M_U^2+m_t^2+{\textstyle{2\over 3}}(m_Z^2-m_W^2)c_{2\beta}
                       \end{array}\right]\nonumber\\ \\
M^2_{\tilde b}&=&\left[\begin{array}{cc}
    M_Q^2+m_b^2-\sixth(2m_W^2+m_Z^2)c_{2\beta}&
    m_b(A_b-\mu t_{\beta})\\
    m_b(A_b-\mu t_{\beta})&
    M_D^2+m_b^2-\third(m_Z^2-m_W^2)c_{2\beta}
                       \end{array}\right].\nonumber
\label{eq:sqmassmat}
\end{eqnarray}
where the squark masses $M_Q$, $M_U$, and $M_D$, the trilinear
couplings $A_t$ and $A_b$, and the supersymmetric Higgs mass
parameter $\mu$ are all defined in section 2.


\section*{Figure Captions}

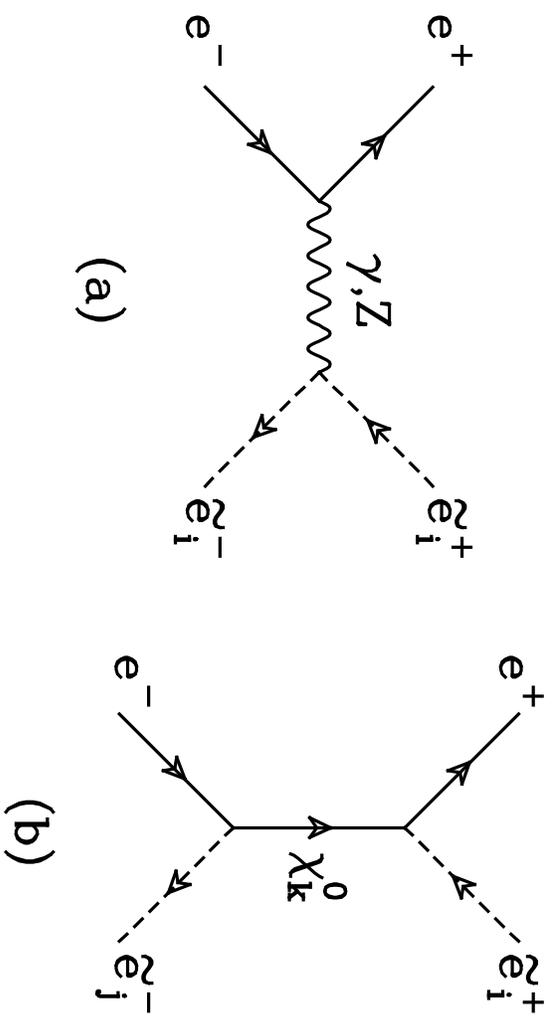
\noindent {\bf Figure 1:}
Feynman diagrams contributing to selectron pair production in
electron--positron annihilation.
Pairs of the same type, $\tilde e^+_L\tilde e^-_L$ or
$\tilde e^+_R\tilde e^-_R$, are produced through diagrams
(a) and (b). On the other hand, pairs of a different type,
$\tilde e^+_L\tilde e^-_R$
or $\tilde e^+_R\tilde e^-_L$, are produced through diagram
(b) only.

\vskip .4cm
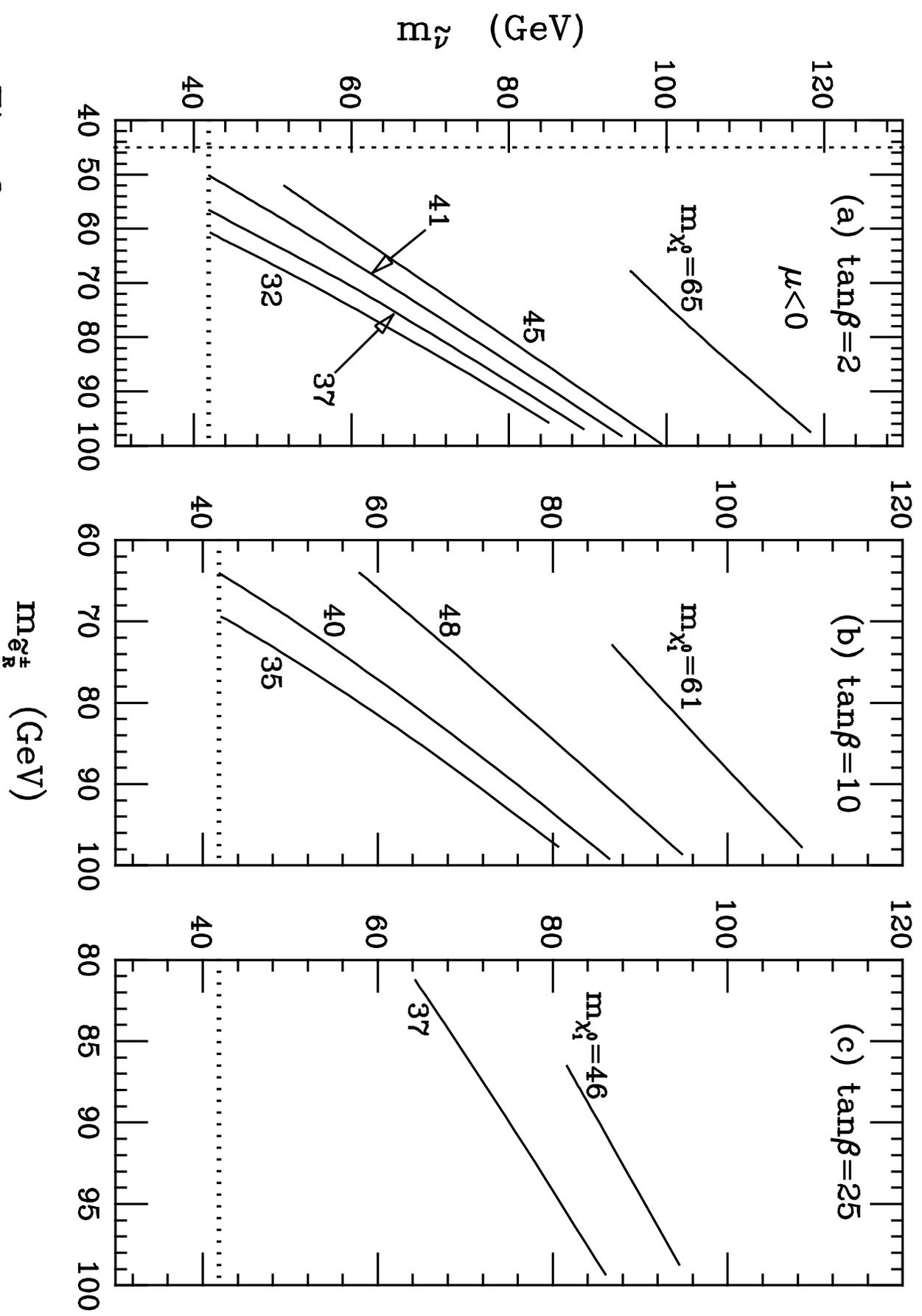
\noindent {\bf Figure 2:}
Sneutrino mass as a function of the lightest selectron mass for
$\mu<0$ and
(a) $\tan\beta=2$, (b) $\tan\beta=10$, and (c) $\tan\beta=25$.
In each case, different curves characterized by a constant value
of the lightest neutralino mass are presented, as indicated in
the figure. The experimental bound $m_{\tilde\nu}>41.8$ GeV
appears as a horizontal dotted line, and the bound
$m_{\tilde e^{\pm}}>45$ GeV is visible only in case (a) as a
vertical dotted line.

\vskip .4cm
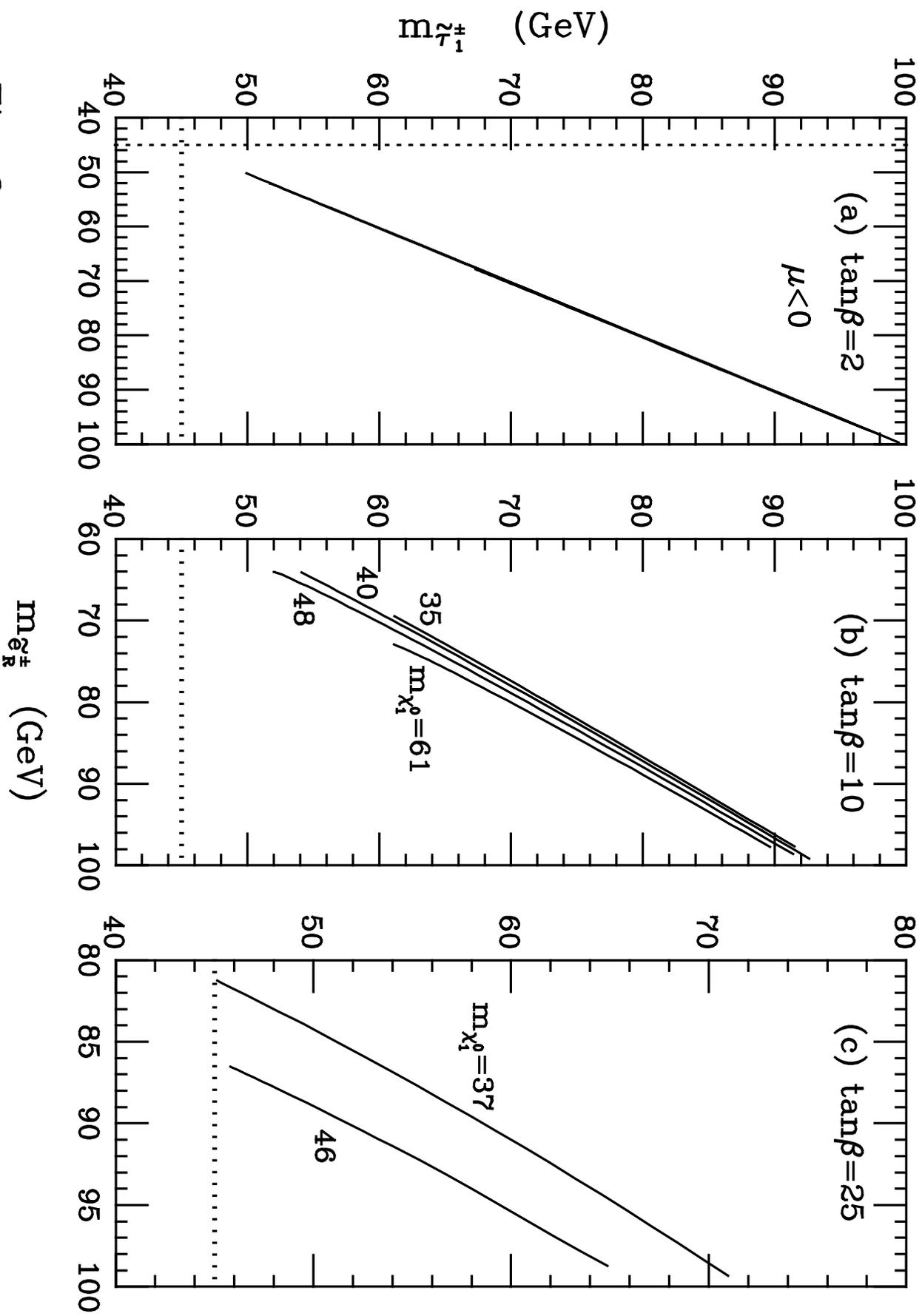
\noindent {\bf Figure 3:}
Lightest stau mass as a function of the $\tilde e^{\pm}_R$ mass
for $\mu<0$ and (a) $\tan\beta=2$, (b) $\tan\beta=10$, and (c)
$\tan\beta=25$.
In each case, the curves are labeled by the same values of the
lightest neutralino mass as in Fig.~2. In case (a), the stau
mass turns out to be independent of the neutralino mass, therefore,
the five curves appear superimposed. The horizontal dotted lines
correspond to the experimental lower bound
$m_{\tilde\tau^{\pm}}>45$ GeV.

\vskip .4cm
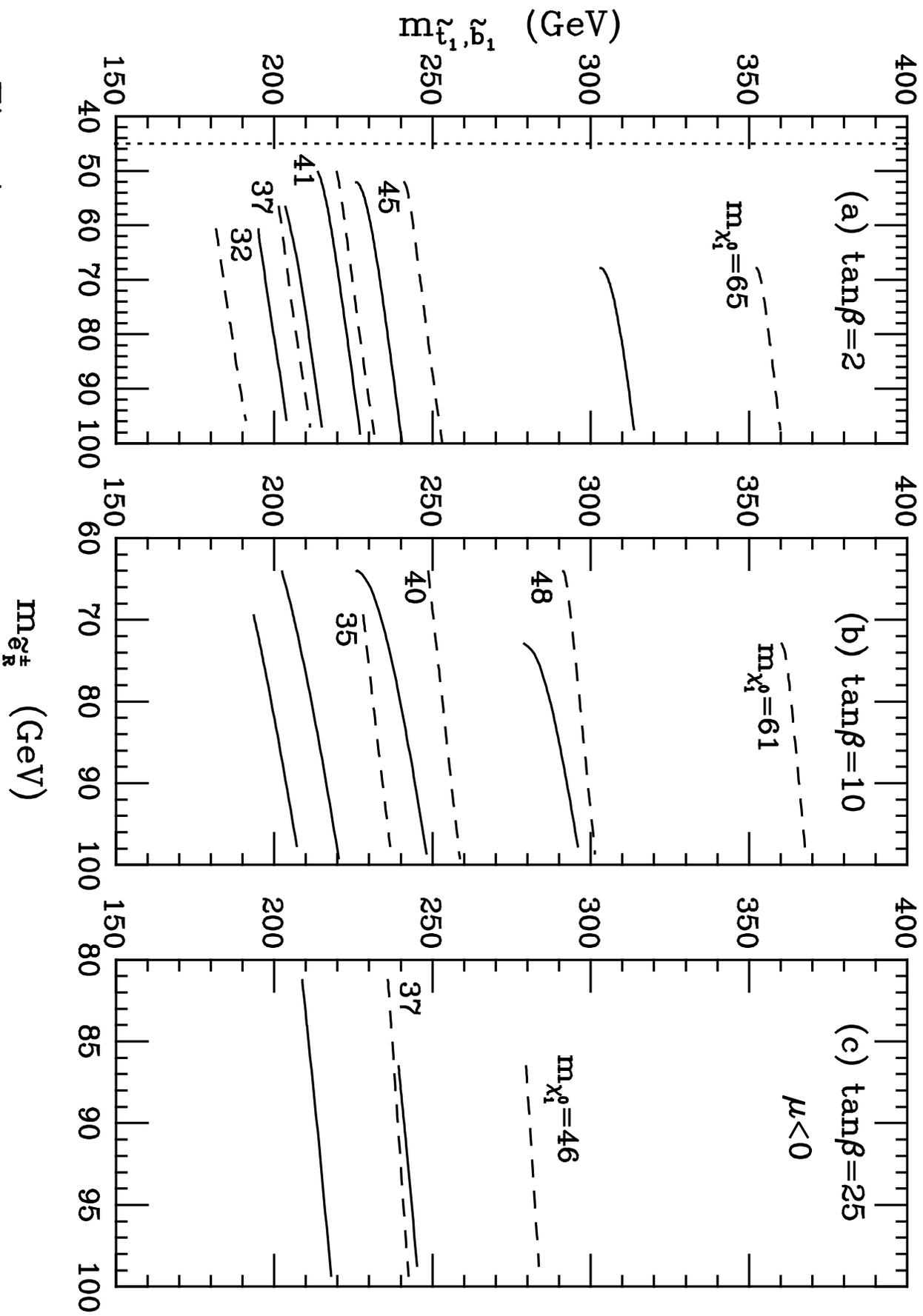
\noindent {\bf Figure 4:}
Mass of the lightest top (solid line) and bottom (dashed line)
squarks as a function of the $\tilde e^{\pm}_R$ mass for $\mu<0$ and
(a) $\tan\beta=2$, (b) $\tan\beta=10$, and (c) $\tan\beta=25$.
The curves are labeled by the value of the $\chi^0_1$ mass as
in Fig.~2.

\vskip .4cm
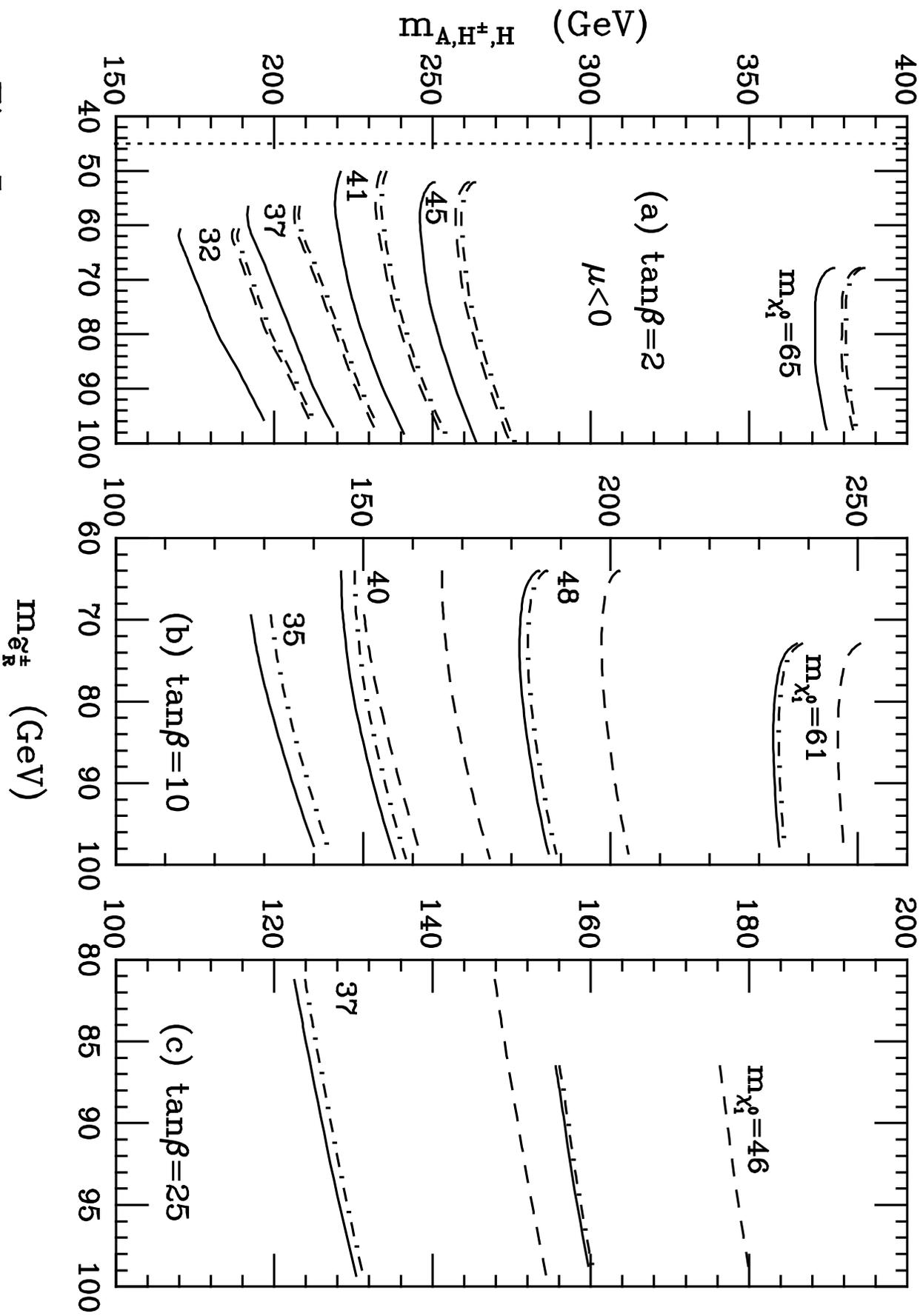
\noindent {\bf Figure 5:}
Mass of the CP-odd (solid), charged (dashes), and heavy CP-even
(dotdashes) Higgs bosons as a function of the $\tilde e^{\pm}_R$
mass for $\mu<0$ and (a) $\tan\beta=2$, (b) $\tan\beta=10$, and (c)
$\tan\beta=25$. The curves are labeled by the value of the
$\chi^0_1$ mass as in Fig.~2.

\vskip .4cm
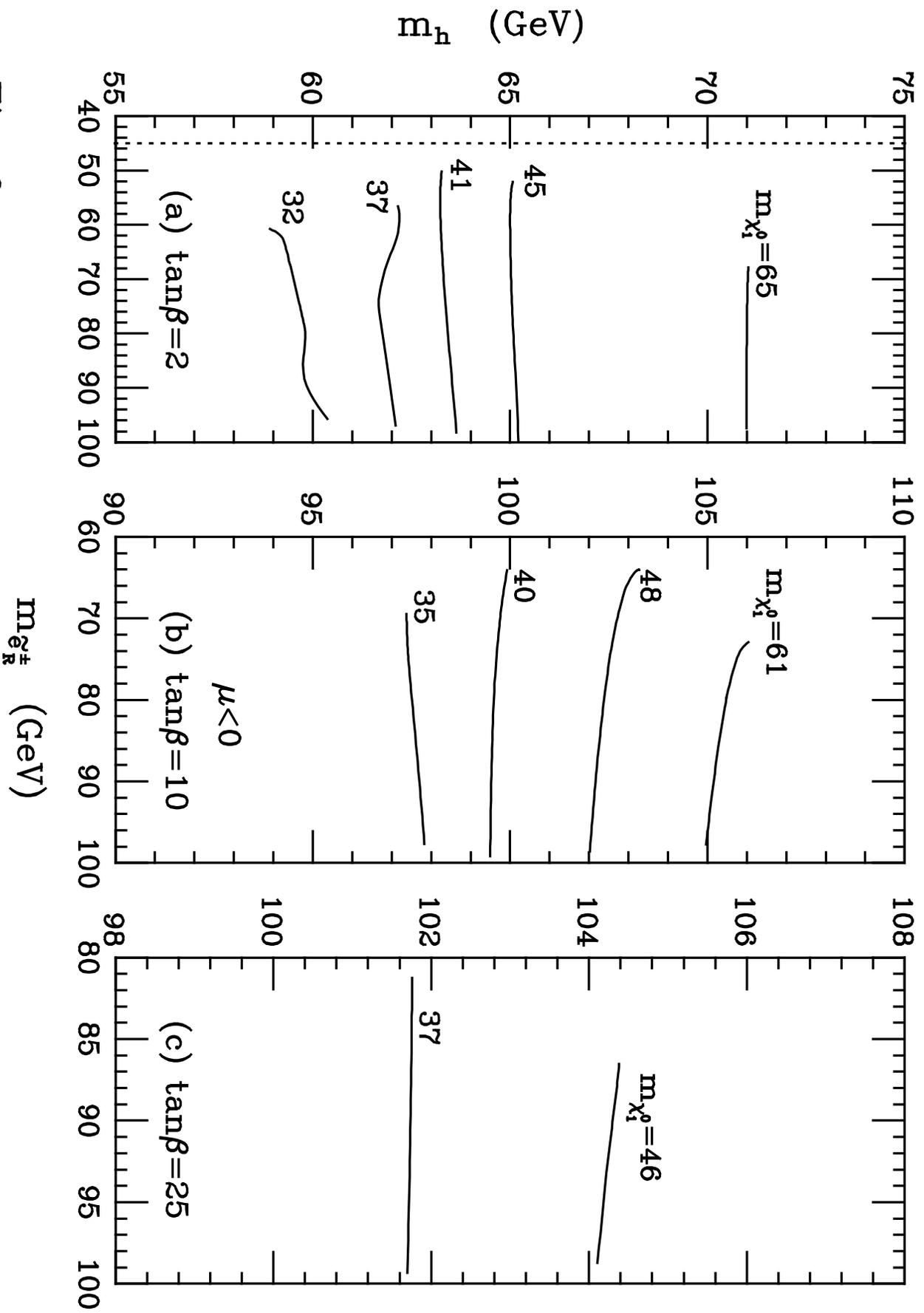
\noindent {\bf Figure 6:}
Lightest CP-even Higgs mass as a function of the $\tilde e^{\pm}_R$
mass for $\mu<0$ and (a) $\tan\beta=2$, (b) $\tan\beta=10$, and (c)
$\tan\beta=25$. The curves are labeled by the value of the
$\chi^0_1$ mass as in Fig.~2.

\vskip .4cm
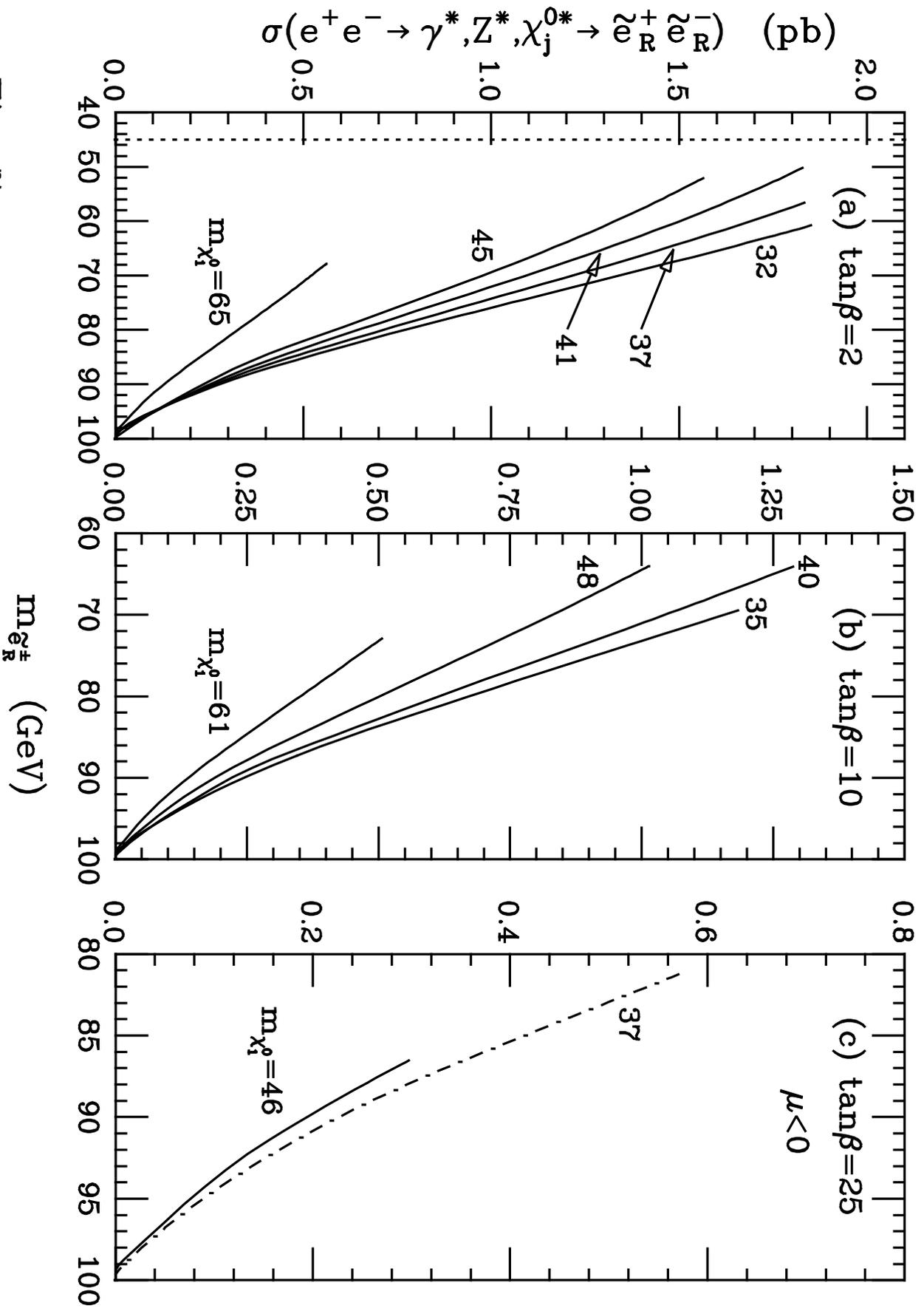
\noindent {\bf Figure 7:}
Total cross section for the production of a pair of right selectrons
in $e^+e^-$ annihilation as a function of the $\tilde e^{\pm}_R$ mass
for $\mu<0$ and (a) $\tan\beta=2$, (b) $\tan\beta=10$, and (c)
$\tan\beta=25$. In case (c) the dashed-dotted line indicates that the
corresponding prediction for the BR($b\rightarrow s,\gamma$) is
excluded. The curves are labeled by the value of the
$\chi^0_1$ mass as in Fig.~2.

\vskip .4cm
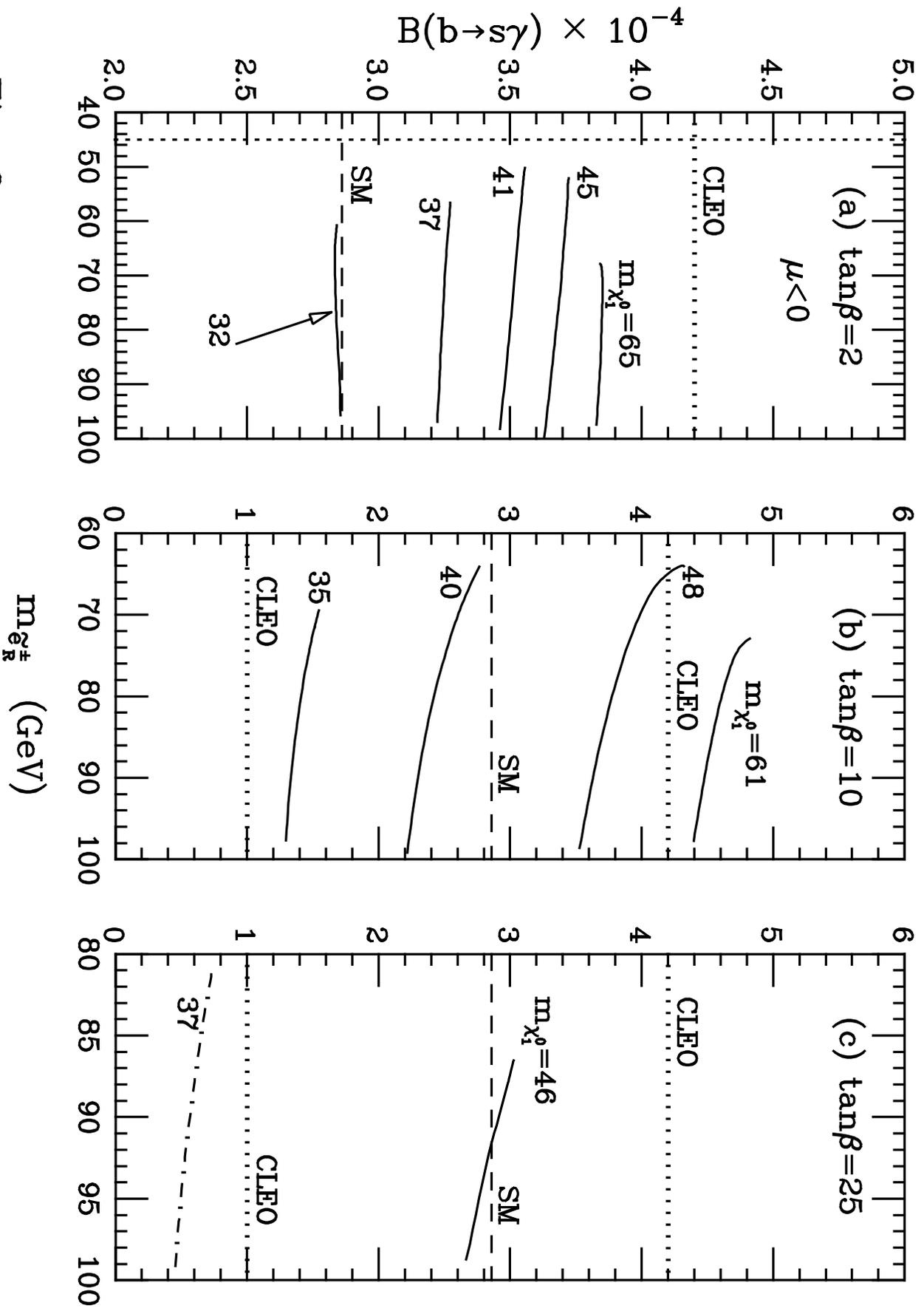
\noindent {\bf Figure 8:}
Inclusive branching ratio of the decay $b\rightarrow s\gamma$
as a function of the $\tilde e^{\pm}_R$ mass for $\mu<0$ and (a)
$\tan\beta=2$, (b) $\tan\beta=10$, and (c) $\tan\beta=25$.
The curves are labeled by the value of the $\chi^0_1$ mass
as in Fig.~2. CLEO upper and lower bounds are shown as
horizontal dotted lines, and the SM prediction appears as
an horizontal dashed line. The dashed--dotted line in case (c)
is excluded even after considering the 25\% theoretical error.

\vskip .4cm
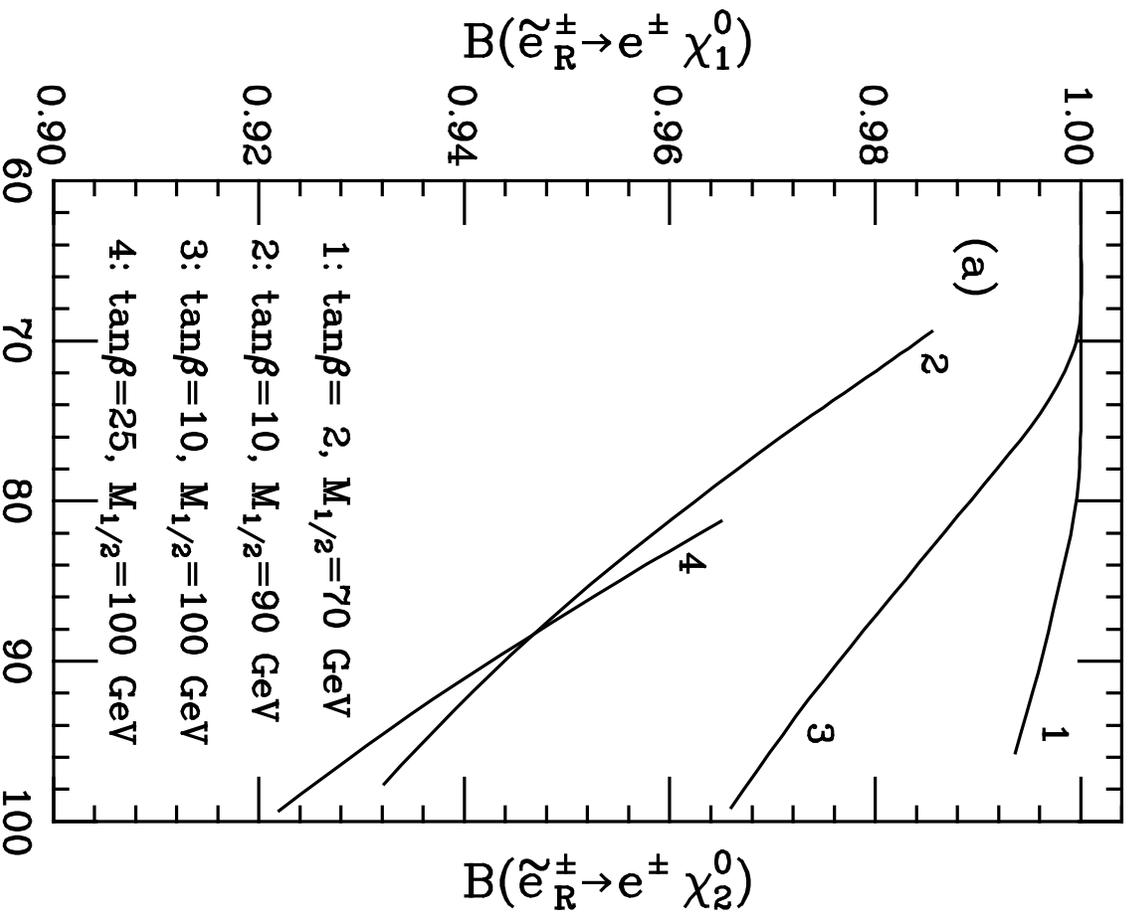
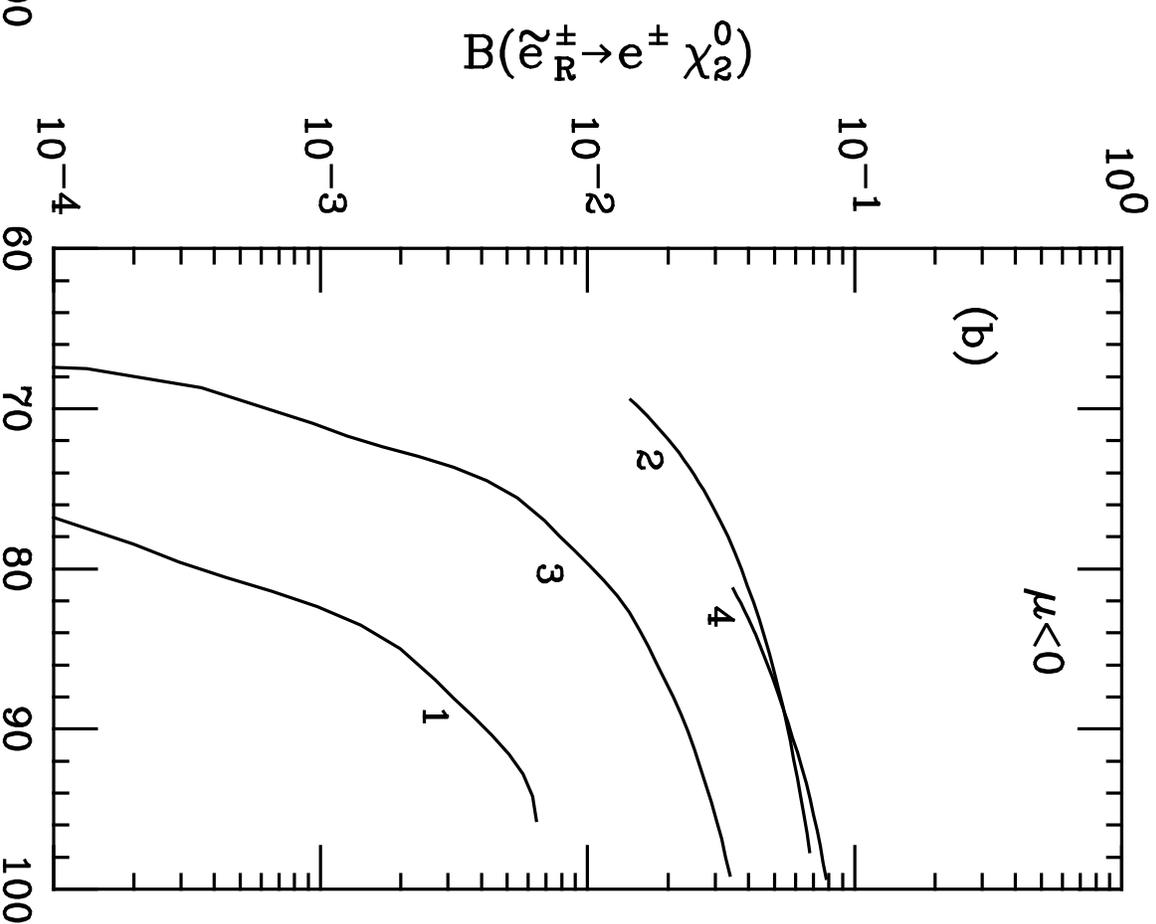
\noindent {\bf Figure 9:}
Branching ratios of the decay of a right selectron into
neutralinos and an electron as a function of $m_{\tilde e^{\pm}_R}$.
We consider $\mu<0$ and take different values of $\tan\beta$
and the universal gaugino mass $M_{1/2}$. Only (a) $\chi^0_1$
and (b) $\chi^0_2$ are light enough for the decay to be allowed.

\vskip .4cm
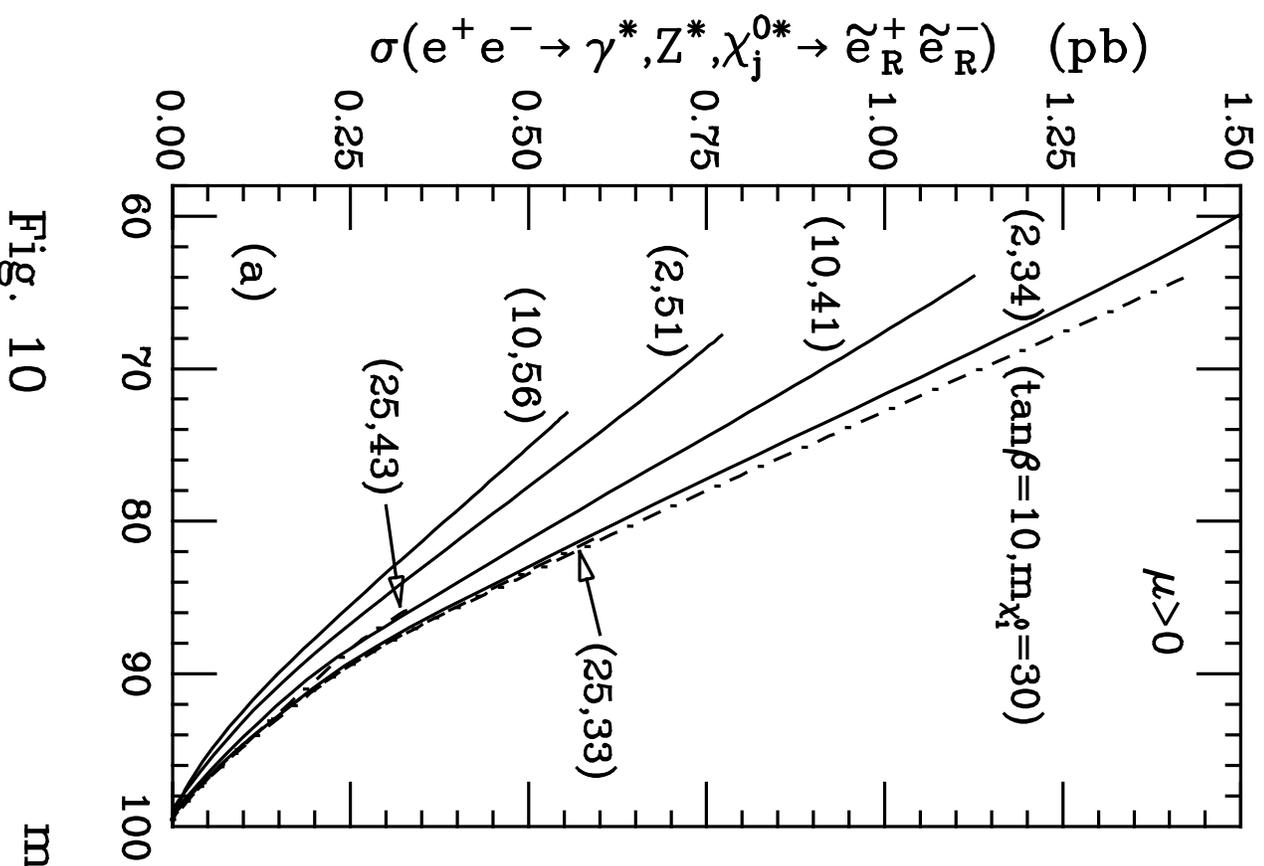
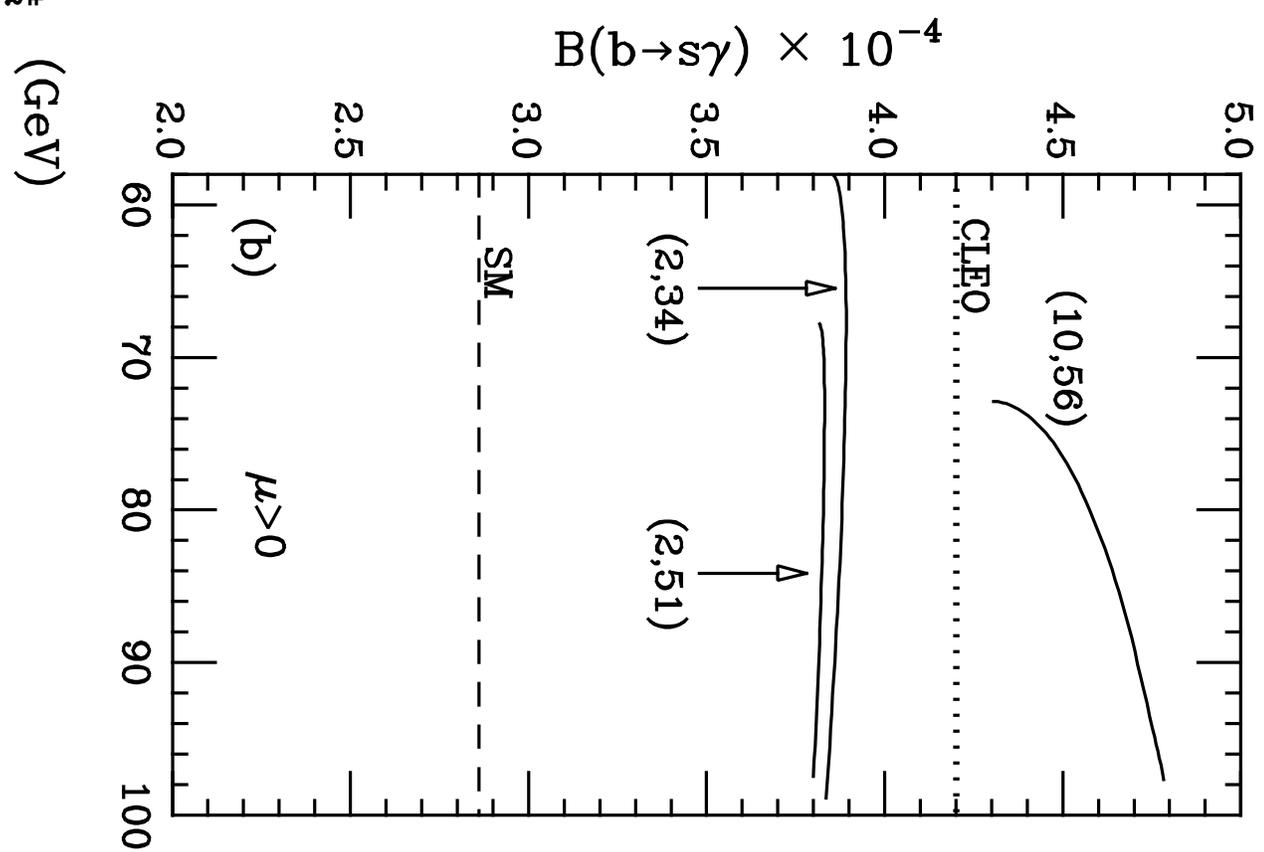
\noindent {\bf Figure 10:}
As a function of the selectron mass, we plot (a) the total cross
section of selectron pair production, and (b) the branching ratio of
the inclusive decay $b\rightarrow s\gamma$, for the case $\mu>0$. Each
curve is labeled by the values of $\tan\beta$ and $m_{\chi^0_1}$. Some
of the solid curves in (a) are not plotted in (b) because their
corresponding prediction for the BR is far above the upper limit we
have set on the y--axis, although still compatible with the
experimental limit if we allow for the $25 \%$ error quoted for the
theoretical calculation. On the contrary, the dashed--dotted lines
are excluded even after considering this error.

\vskip .4cm
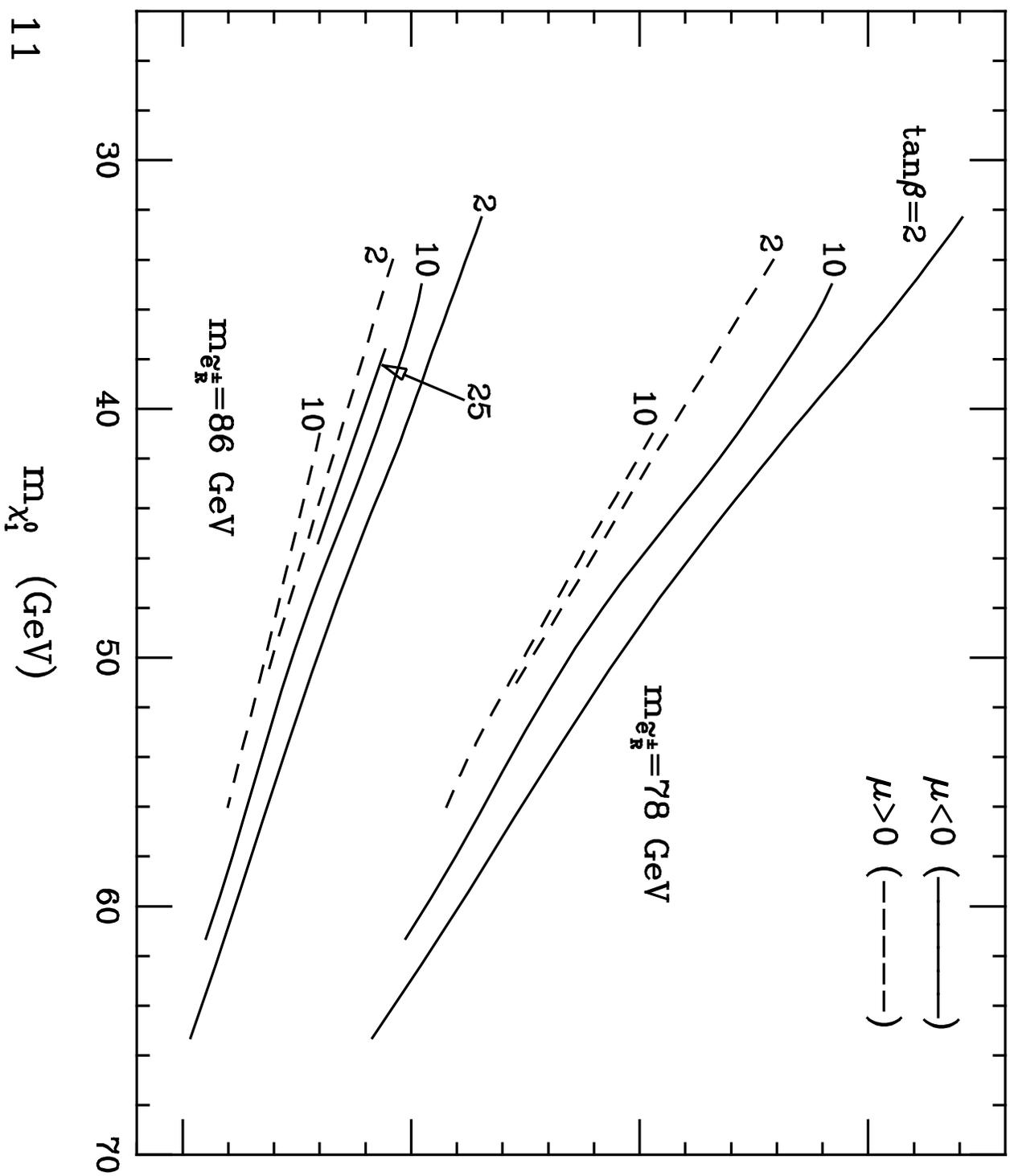
\noindent {\bf Figure 11:}
Total cross section for the production of a pair of right
selectrons in $e^+e^-$ annihilation as a function of the
lightest neutralino mass. The curves are defined by a
constant value of the right selectron mass and a constant
value of $\tan\beta$ as indicated by the figure. Solid lines
correspond to $\mu<0$ while dashed lines correspond to $\mu>0$.

\end{document}